\documentclass[prd,10pt,twocolumn,tightenlines,superscriptaddress,numerical,notitlepage,showpacs,amsmath,amssymb,amsfonts,aps,nofootinbib,longbibliography,floatfix]{revtex4-1}

\usepackage{combelow}
\usepackage{dsfont}
\usepackage{mathtools,bm,bbm,braket,slashed}
\usepackage{tikz}

\usepackage{xcolor}
\usepackage{xurl}
\usepackage[breaklinks=true]{hyperref}
\hypersetup{colorlinks=true, citecolor=blue, urlcolor=blue, citecolor=red}

\usepackage{MnSymbol,wasysym}
\usepackage{tikz-feynman}
\tikzfeynmanset{compat=1.1.0}

\newcommand{\varphir}{\varphi_\mathrm{r}}
\newcommand{\bp}{\bm{p}}
\newcommand{\br}{\bm{r}}
\newcommand{\bx}{\bm{x}}
\newcommand{\bJ}{\bm{J}}
\newcommand{\calF}{\mathcal{F}}
\newcommand{\calJ}{\mathcal{J}}
\newcommand{\calI}{\mathcal{I}}
\newcommand{\bcalJ}{\bm{\mathcal{J}}}
\newcommand{\bnabla}{\bm{\nabla}}
\newcommand{\bOmega}{\bm{\Omega}}
\newcommand{\modep}{\tilde{\varepsilon}}

\newcommand{\propx}{G} 
\newcommand{\propp}{\widetilde{G}} 

\newcommand{\selfp}{\widetilde{\Pi}}
\newcommand{\nb}{n_{\mathrm{B}}}

\begin{document}

\title{Moment of Inertia of an Interacting Bose Gas}

\date{\today}

\author{Aleksandar Geci\'c}
\affiliation{Department of Physics, West University of Timi\cb{s}oara, Bd.~Vasile P\^arvan 4, Timi\cb{s}oara 300223, Romania}

\author{Victor E. Ambru\cb{s}}
\affiliation{Department of Physics, West University of Timi\cb{s}oara, Bd.~Vasile P\^arvan 4, Timi\cb{s}oara 300223, Romania}

\author{Kenji Fukushima}
\affiliation{Department of Physics, The University of Tokyo,\\
7-3-1 Hongo, Bunkyo-ku, Tokyo 113-0033, Japan}
\affiliation{Department of Physics, West University of Timi\cb{s}oara, Bd.~Vasile P\^arvan 4, Timi\cb{s}oara 300223, Romania}

\begin{abstract}
The response of many-body quantum systems to rotation can be characterized by the moment of inertia. For a classical
gas, the moment of inertia can be expressed as the integral of the enthalpy density multiplied by the squared radial distance from the rotation axis.
In quantum field theory, the finite rotation in the grand canonical ensemble demands the causality bound. This constraint imposes technical challenges in treating transverse momenta discretized with the Bessel function zeros.
However, we define the moment of inertia in the limit of zero angular velocity, in which the causality constraint is irrelevant and ordinary quantum field theoretical techniques can be applied.  We evaluate the moment of inertia in the $\phi^4$ theory and find that, surprisingly, the interacting effects including the ring-diagram resummation are consistent with the classical expectation and the moment of inertia density remains proportional to the enthalpy density.
\end{abstract}

\maketitle

\section{Introduction}
\label{sec:intro}

An important probe of the quark-gluon plasma (QGP), 
the state of strongly interacting matter at extreme temperatures and densities, which is produced in ultrarelativistic ion collisions, emerged with the STAR collaboration's observation of $\Lambda$ and $\bar{\Lambda}$ spin polarization \cite{STAR:2017ckg}. The observation of non-vanishing spin polarization implied that the produced medium initially exhibited a non-trivial vortical structure inheritated from the large angular momentum retained in the QGP in non-central collisions. This indicates that a significant portion of the initial orbital momentum is transmitted to the spin degrees of freedom through spin-orbital coupling, a quantum phenomenon inherent in relativistic field theory \cite{Itzykson:1980rh,Peskin:1995ev}. The purpose of our present investigation is to establish and examine the system's response to the rotation, by studying the moment of inertia of an interacting quantum system. 

A non-perturbative assessment can be obtained from first principles using the lattice QCD approach, which indicates that the moment of inertia can become negative above the QCD transition temperature $T_c$, up to the supervortical temperature, $T_{sv} \simeq 1.5 T_c$ \cite{Braguta:2023yjn, Braguta:2024zpi, Braguta:2023tqz, Braguta:2025yud}. This negative contribution was speculated to originate from the evaporation of the nonpertubrative chromomagnetic component and its coupling to the angular velocity \cite{Braguta:2023tqz}. Due to the infamous sign problem, these calculations were performed in Euclidean space for imaginary rotation parameter, $\Omega = i \Omega_I$, with $\Omega_I$ a real number. An important result obtained using lattice QCD, possibly linked to the emergence of a negative moment of inertia, is the decrease of the deconfinement transition temperature as the {\it imaginary} angular velocity is increased. Analytic continuation to real angular velocities implies an {\it increase} of the deconfinement transition temperature as the {\it real} angular velocity is increased.

Analyses based on several effective models in the mean-field approximation show no indication of a negative moment of inertia. Studies employing the Nambu-Jona-Lasinio (NJL) model \cite{Jiang:2016wvv,Chernodub:2016kxh,Chernodub:2017ref,Wang:2018sur, Kawaguchi:2025mkh}, linear sigma model (LSM) \cite{Chen:2023cjt,Hernandez:2024nev}, their Polyakov loop extensions \cite{Singha:2024tpo,Singha:2025zvh}, hadron resonance gas \cite{Fujimoto:2021xix} and perturbative QCD \cite{Chen:2022smf,Chen:2024tkr} display a clear disagreement with the lattice QCD results. Similar trend was found by holographic QCD approach \cite{Braga:2022yfe, Chen:2020ath}.

A possible way to achieve agreement with lattice results was proposed in Ref.~\cite{Sun:2024anu}, inspired from the work of incorporating the inverse magnetic catalysis effect into effective models, originally proposed by Farias et al~\cite{Farias:2014eca}, where agreement with lattice data was obtained by making the NJL coupling parameter $G$ a function of the magnetic field. Similarly, turning the coefficients of the Polaykov loop effective potential into angular velocity-dependent functions allow the effective model predictions to reproduce lattice data, hence leading to an agreement that the deconfinement temperature increases as the angular velocity is increased.

Another possible origin of the negative moment of inertia can be related to the nonanalytic structures leading to the fractal thermodynamics in systems under imaginary rotation \cite{Ambrus:2023bid,Patuleanu:2025zbn}. Ref.~\cite{Ambrus:2023bid} considering the case of a rotating scalar field shows that the moment of inertia under imaginary rotation $\Omega = i \Omega_I$ can alternate between positive and negative values, depending on the dimensionless number $\nu = \beta \Omega_I / 2\pi$ and on the size of the system under consideration. This conclusion seems compatible with the equivalent fermion system examined in Ref.~\cite{Patuleanu:2025zbn}, however the moment of inertia was not reported there.

Yet another possible origin for the lattice QCD negative moment of inertia is the gluon self-interactions, which eventually lead to the buildup of the chromomagnetic condensates. A recent analysis by Siri and Sadooghi \cite{Siri:2024scq} of the self-interacting $\phi^4$ scalar field theory indicate that the nonanalytic $\lambda^{3/2}$ contribution to the thermodynamic potential, arising from the resummation of infrared-divergent ring diagrams, makes negative contributions to the moment of inertia, eventually leading to a negative overall value at large enough values of the coupling parameter.

In this work, we reanalyze the self-interaction contributions in the simplified $\phi^4$ theory. While this toy model is still far from the complicated structure of the $SU(3)$ Yang-Mills theory, our purpose is to elucidate the role of self interactions in the response of the field's mechanical properties with respect to rotation. We question the validity of the results reported in Ref.~\cite{Siri:2024scq}, which were obtained using an integration over the whole space of the thermodynamic potential under rotation, expressed with respect to cylindrical modes. We challenge this procedure, as the integration over an infinite volume and setting $\Omega = 0$ are mathematical operations which may not commute. We provide an alternate derivation of the moment of inertia in this system by first deriving the expression for the moment of inertia and its local point-dependent density within the theory under rotation, while performing the loop computations only in the limit of vanishing rotation. In our calculation, we rely on standard thermal field theory techniques that are well-established for static states \cite{Kapusta:2007xjq}. Our calculations yield a result in disagreement with Ref.~\cite{Siri:2024scq}, namely that the moment of inertia density is simply $\mathcal{I} = r_\perp^2 h$, with $r_\perp$ being the perpendicular distance to the rotation axis and $h = p + e$ the enthalpy density. This conclusion is supported by our calculations in the case of the Dirac field~\cite{Patuleanu:2025zbn,Singha:2025zvh} (note that in the Dirac field case, there is also a spin contribution).

Here, we mention that another approach to consistently include non-vanishing rotation is to enclose the system inside a cylindrical boundary placed within the light cylinder, $R \Omega < 1$. This procedure cuts out the spatial region outside the light cylinder, where thermal expectation values are undefined \cite{Duffy:2002ss,Ambrus:2015lfr}. This approach was outlined by Kuboniwa and Mameda in Ref.~\cite{Kuboniwa:2025vpg}, where the authors point out the importance of keeping track of the point-dependent scalar field fluctuations, leading to a point dependence of the field self-energy. Technical complications related to the evaluation of the point-dependent self-energy kernel by summing over Bessel functions make this approach significantly more challenging and an evaluation of the moment of inertia in this setup is still absent \cite{Kuboniwa:2025vpg}.

This paper is structured as follows. In Sec.~\ref{sec:gen}, we formulate the moment of inertia and its local density within field theory, using the path-integral formalism for rotating states. In Sec.~\ref{sec:heuristic}, we provide heuristic derivations of the moment of inertia within the classical kinetic theory approach and using a quantum field-theoretical formulation of the partition function with cylindrical modes, thereby establishing our result: $\mathcal{I} = r_\perp^2 h$. Section~\ref{sec:loop} presents the detailed loop computation within the framework of nonrotating thermal field theory, providing an explicit link between the moment of inertia and the two- and four-point functions of the interacting theory. A consistent derivation of the enthalpy density is discussed in Sec.~\ref{sec:loop:h}. Our conclusions are presented in Sec.~\ref{sec:conc}. 

\section{Rotating systems: general considerations}\label{sec:gen}

In this section, we lay the foundations for our study of the moment of inertia in quantum systems under rotation. In Subsec.~\ref{sec:gen:gce}, we derive the expression for the moment of inertia at vanishing rotation in the grand canonical ensemble. In Subsec.~\ref{sec:gen:qft}, we derive the expression for the moment of inertia in the static limit for the self-interacting scalar field.

\subsection{Grand canonical ensemble description}
\label{sec:gen:gce}

Thermal properties in a physical system under rigid rotation with the angular velocity vector $\bOmega$ at temperature $T = \beta^{-1}$ are described in the grand canonical ensemble by the partition function and the density matrix,
\begin{equation}
  Z = \mathrm{tr}\hat{\rho}\,,
  \qquad
  \hat{\rho} = e^{-\beta (\hat{H} - \bOmega \cdot \hat{\bJ})} \,,
\end{equation}
where $\hat{H}$ and $\hat{\bm{J}}$ are the Hamiltonian and the total angular momentum operators.  The free energy $F = -T \ln Z$ satisfies the following thermodynamic relation:
\begin{equation}
  dF = -S dT - p dV - \bJ \cdot d\bOmega
\end{equation}
with the total entropy $S$, the pressure $p$ and the total angular momentum expectation value $\bJ$ given by
\begin{equation}
  S = -\frac{\partial F}{\partial T}\,,
  \qquad
  p = -\frac{\partial F}{\partial V}\,,
  \qquad
  \bJ = -\frac{\partial F}{\partial \bOmega} \,.
\end{equation}
Due to centrifugal effects, the rotating state is inhomogeneous and local expectation values (e.g., energy-momentum tensor and angular momentum) develop a dependence on the cylindrical radial distance, $r_\perp = |\br_\perp|$.
These expectation values may have a singularity at the causality bound, i.e., $r_\perp\Omega \to 1$, where $\Omega=|\bOmega|$.

It is convenient to introduce the free energy density in such a way as to satisfy:
\begin{equation}
    F = \int d^3x\, \calF(\bx)\,.
\end{equation}
From this expression, the integrand $\calF(\bx)$ is not uniquely determined in general; we can add total derivative terms as long as the surface terms do not change the integral.  As we will see later, however, we are looking for the integrand $\propto r_\perp^2$, and this type of integrand has no ambiguity because surface terms are nonzero at the boundary.
We can define the angular momentum density, $\bcalJ(\bx)$, by introducing a local vorticity, $\bOmega(\bx)$, or using the free energy density, $\calF(\bx)$, as
\begin{equation}
  \bcalJ(\bx) = -\frac{\delta F}{\delta\bOmega(\bx)} = -\frac{d\calF(\bx)}{d\bOmega} \,,
\end{equation}
which is consistent with $\bJ = \int d^3x\,\bcalJ(\bx)$.

In this paper, we treat the rigid rotation along the $z$ axis such that $\bJ = J \bm{e}_z$, $\bcalJ = \calJ \bm{e}_z$, and $\bOmega = \Omega \bm{e}_z$.  We then consider only the diagonal component of the moment of inertia tensor as\footnote{We could keep $\Omega$ finite after taking the derivative, but such a quantity is usually called the rotational susceptibility.}
\begin{equation}
  I = \frac{d J}{d\Omega} \biggr\rvert_{\Omega=0} \,. 
  \label{eq:I_def}
\end{equation}
In the same way as we defined $\calF(\bx)$, we can introduce the moment of inertia density, $\calI(\bx)$, so that $I = \int d^3x\, \calI(\bx)$ can hold.  Clearly, we can deduce $\calI(\bx)$ from the following relations:
\begin{equation}
  \calI(\bx) = \frac{\delta J}{\delta \Omega(\bx)} \biggr\rvert_{\Omega=0} = \frac{\partial \calJ(\bx)}{\partial \Omega}\biggr\rvert_{\Omega=0} = - \frac{\partial^2 \calF(\bx)}{\partial \Omega^2}\biggr\rvert_{\Omega = 0} \,.
\end{equation}
Here, we explicitly see that the moment of inertia (density) is a well-defined thermodynamic quantity in the ordinary framework of quantum field theory at $\Omega = 0$.

\subsection{Scalar field theory with rotation}
\label{sec:gen:qft}

The scalar field theory with self-interactions is described by the following Lagrangian density: 
\begin{equation}
  \mathcal{L} = \frac{1}{2} \partial_\mu \phi \partial^\mu \phi - \frac{1}{2} m^2 \phi^2 - V(\phi) \,,
  \quad 
  V(\phi) = \frac{\lambda}{4!} \phi^4 \,.
 \label{eq:L}
\end{equation}
Switching to the co-rotating coordinates defined by
\begin{equation}
  t_\mathrm{r} = t \,,
  \qquad 
  \varphi_\mathrm{r} = \varphi - \Omega t \,,
\end{equation}
we rewrite the Lagrangian density as
\begin{equation}
 \mathcal{L} = \frac{1}{2} (\partial_{t_\mathrm{r}} \phi - \Omega \partial_{\varphi_\mathrm{r}} \phi)^2 - \frac{1}{2} (\bnabla_\mathrm{r} \phi)^2 - \frac{1}{2} m^2 \phi^2 - V(\phi) \,.
 \label{eq:Lrot}
\end{equation}
According to the power in $\Omega$, we reorganize this Lagrangian density as
\begin{equation}
 \mathcal{L} = \mathcal{L}_{(0)} + \Omega \mathcal{L}_{(1)} + \frac{\Omega^2}{2} \mathcal{L}_{(2)} \,,
\end{equation}
where
\begin{subequations}
  \begin{align}
    \mathcal{L}_{(0)} &= \frac{1}{2} \left[(\partial_{t_{\rm r}} \phi)^2 - (\bnabla_{\rm r}\phi)^2 \right] - V(\phi) \,,\\
    \mathcal{L}_{(1)} &= -\partial_{t_\mathrm{r}} \phi \partial_{\varphir} \phi \,,\\
    \mathcal{L}_{(2)} &= (\partial_{\varphir} \phi)^2 \,.
  \end{align}
\end{subequations}
It should be noted that we can rewrite $\mathcal{L}_{(1)} = \mathcal{M}^z = \mathcal{M}^{0,xy}$ where the angular momentum density operator, $\mathcal{M}^{\lambda,\mu\nu}$, and the energy-momentum tensor operator, $\Theta^{\mu\nu}$, are defined, respectively, as
\begin{align}
  \mathcal{M}^{\lambda,\mu\nu} &= x^\mu \Theta^{\lambda\nu} - x^\nu \Theta^{\lambda\mu} \,,\\
  \Theta^{\mu\nu} &= \partial^\mu \phi \partial^\nu \phi - g^{\mu\nu} \mathcal{L} \,.
  \label{eq:M_Theta}
\end{align}

In the Wick rotated Euclidean coordinates with replacement of $\mathcal{L}_\mathrm{E}(\tau_\mathrm{r}, \bx_\mathrm{r}) = -\mathcal{L}(t_\mathrm{r} \to -i\tau_\mathrm{r}, \bx_\mathrm{r})$, we can compute thermodynamic quantities from the partition function,
\begin{equation}
  Z = \int [d\phi]\, e^{-S_\mathrm{E}} \,, \qquad 
  S_\mathrm{E} = \int d^4 X_\mathrm{r}\, \mathcal{L}_\mathrm{E} \,.
\end{equation}
We introduced a short-hand notation, $d^4 X_\mathrm{r}=d\tau_\mathrm{r} d^3 x_\mathrm{r}$ where the $\tau_\mathrm{r}$ integration ranges in $[0,\beta)$ with the periodic boundary condition, $\phi(\tau_\mathrm{r} + \beta, \bx_\mathrm{r}) = \phi(\tau_\mathrm{r}, \bx_\mathrm{r})$.  In terms of $Z$, the quantities of our interest take the following expressions:
\begin{equation}
  J = \frac{T}{Z} \frac{\partial Z}{\partial \Omega} \,,
  \qquad 
  I = \frac{\partial J}{\partial \Omega} \biggr\rvert_{\Omega = 0} = \frac{T}{Z} \frac{\partial^2 Z}{\partial \Omega^2} \biggr\rvert_{\Omega=0}\,.
\end{equation}
In the expression for $I$, the first derivative does not appear because $J=0$ in the $\Omega\to 0$ limit.

From the above definitions, we can take the second derivative and split the results into two parts, $I = I_1 + I_2$, where
\begin{align}
  I_1 &= T \int_{X_1,X_2} \langle \mathcal{L}_{(1)}(X_1) \mathcal{L}_{(1)}(X_2) \rangle
  = T \int_{X_1,X_2} \langle \mathcal{M}_1^z \mathcal{M}_2^z \rangle \,,\nonumber\\
  I_2 &= T \int_{X} \langle \mathcal{L}_{(2)}(X) \rangle\,.\label{eq:I1I2}
\end{align}
Here, $\int_X = \int d^4X$
and $\int_{X_1, X_2} = \int d^4 X_1 \int d^4 X_2$, for notational brevity.

For later convenience, let us rewrite the above integrals, $I_1$ and $I_2$, using standard correlation functions in quantum field theory.
Introducing the auxiliary variables, $X_1'$, $X_2'$ for $I_1$ and $X'$ for $I_2$, they are built with the thermal two-point and four-point functions as follows:
\begin{align} 
  I_1 &= T \int_{X_1,X_1',X_2,X_2'} \delta^{(4)}(X_1 - X_1') \delta^{(4)}(X_2 - X_2') \nonumber\\
 &\qquad\qquad \times \partial_{t_1}  \partial_{\varphi_1'} \partial_{t_2} \partial_{\varphi_2'} G_4(X_1, X_1', X_2, X_2') \,,
  \label{eq:I1} \\
  I_2 &= T \int_{X,X'} \delta^{(4)}(X - X') \partial_\varphi \partial_{\varphi'} G_2(X, X') \,,
  \label{eq:I2}
\end{align}
where
\begin{align}
 G_2(X, X') &= \langle \hat{\phi}(X) \hat{\phi}(X') \rangle \,, \\
 G_4(X_1, X_1', X_2, X_2') &= \langle \hat{\phi}(X_1) \hat{\phi}(X_1') \hat{\phi}(X_2) \hat{\phi}(X_2') \rangle\,.
\end{align}
It is important to emphasize that these expectation values are taken at vanishing rotation; thereby, for the evaluation of the moment of inertia, we can safely employ standard thermal field theory techniques.  Nevertheless, as we will see later, the phase space integrations involve nonstandard complications.

\section{Heuristic Discussions}\label{sec:heuristic}

We aim to develop a field-theoretical framework and proceed with diagrammatic calculations in detail.  Prior to technical discussions, it should be useful to make our claim clear first and then provide some heuristic arguments to justify it.

Our calculations will confirm that the moment of inertia density, $\calI$, is related to the enthalpy density, $h=p+e$, where $e$ is the energy density, through a simple relation:
\begin{equation}
    \calI = r_\perp^2\, h \,.
    \label{eq:claim}
\end{equation}
This is a very natural relativistic extension.  In classical physics, the integrand of the moment of inertia is given by the mass density weighted by $r_\perp^2$.  If we see an expression of the ideal relativistic hydrodynamics, $\Theta^{\mu\nu}= h \,u^\mu u^\nu - p g^{\mu\nu}$ with the fluid velocity $u^\mu$, the momentum (energy) flux is identified as $\Theta^{0i} = h\, u^0 u^i$; therefore, the fluid counterpart of the mass density is nothing but the enthalpy density.

Although Eq.~\eqref{eq:claim} was conjectured within the hadron resnance gas model~\cite{Fujimoto:2021xix}, it is a nontrivial question whether Eq.~\eqref{eq:claim} holds in physical systems with interactions.  We will show that Eq.~\eqref{eq:claim} is satisfied for a weakly interacting Bose gas even including a nonanalytical contribution from the ring-diagram resummation.

In Subsec.~\ref{sec:heuristic:rkt}, we derive the moment of inertia using the classical kinetic theory description of a Bose gas.  In Subsec.~\ref{sec:heuristic:thermo}, we consider the free energy of a Bose gas under rotation using thermodynamic quantities based on the 2PI (particle irreducible) formalism. Subsections~\ref{sec:heuristic:lam0} and \ref{sec:heuristic:lam1} present the derivation of moment of inertia for the noninteracting part of $O(\lambda^0)$ and higher-order corrections up to $O(\lambda^{3/2})$, respectively, providing a heuristic derivation of our claim in Eq.~\eqref{eq:claim}.

\subsection{Classical thermal gas described by relativistic kinetic theory}
\label{sec:heuristic:rkt}

As a warmup exercise, we shall compute the moment of inertia within the framework of the collisionless kinetic theory.  Since the kinetic theory assumes a quasi-particle treatment, the system should be classical, but we will consider a thermal distribution with quantum statistics.

Global equilibrium under rotation corresponds to the case when the inverse temperature four-vector $\beta^\mu$
is a Killing vector satisfying $\nabla_\mu \beta_\nu + \nabla_\nu \beta_\mu = 0$.  For rigid rotation around the $z$ axis, the relevant timelike Killing vector field is given by
\begin{equation}
  \beta^\mu \partial_\mu = \beta (\partial_t + \Omega \partial_\varphi)\,,
\end{equation}
which is also consistent with the Lagrangian density in the co-rotating coordinates in Eq.~\eqref{eq:Lrot}.
For $\beta^\mu = \beta_\perp u^\mu$, the local inverse temperature $\beta_\perp$ and the local normalized fluid velocity $u^\mu$ can be extracted as
\begin{equation}
  \beta_\perp = \frac{\beta}{\Gamma_\perp} \,,
  \qquad
  u^\mu = \Gamma_\perp (1, \boldsymbol{\Omega} \times \bm{x})
\end{equation}
with $\Gamma_\perp$ representing the Lorentz factor at the cylindrical radial distance $r_\perp$, i.e.,
\begin{equation}
  \Gamma_\perp = \frac{1}{\sqrt{1 - r_\perp^2 \Omega^2}} \,.
\end{equation}
Thus, $\beta$ corresponds to the inverse temperature on the rotation axis at $r_\perp = 0$ where $\Gamma_\perp = 1$. 

The free energy density of a noninteracting Bose gas reads:
\begin{equation}
  \calF = T \int \frac{d^3p}{(2\pi)^3} \, \ln\bigl(1 - e^{-\beta_\perp p^\mu u_\mu}\bigr) \,,
\end{equation}
where $\beta_\perp p^\mu u_\mu$ follows from the relativistic kinetic theory description 
with four-momentum, $p^\mu=(\varepsilon,\bp)$, where $\varepsilon=\sqrt{\bp^2+m^2}$.
We note that $\beta_\perp p^\mu u_\mu = \beta[\varepsilon - \boldsymbol{\Omega} \cdot (\bm{x}\times \bm{p})]$ completely agrees with the field-theoretical treatment of rotation in which the Hamiltonian is shifted as $H\to H - \boldsymbol{\Omega} \cdot \bm{J}$ as emphasized in Ref.~\cite{Chen:2015hfc}.

It is a feasible program to expand the above expression in terms of $\Omega$ and deduce the moment of inertia.  Here, let us take a different strategy to deepen our insight;  we change the variable $p$ to $p'$ boosted to the local rest frame.  The phase-space integration measure, $d^3 p$ is not Lorentz invariant, while the following combination,
\begin{equation}
    dP = \frac{d^3 p}{(2\pi)^3 \varepsilon} \,,
\end{equation}
is an invariant measure.  This means that $d^3 p$ should transform like $p^0 \sim \varepsilon$ under the Lorentz boost.

In the free energy expression, $p\cdot u = p^\mu u_\mu$ is Lorentz invariant such that $p\cdot u = p^{\prime 0} = \varepsilon'$ in the local rest frame.  From the measure, we have a factor, $p^0 = \Gamma_\perp [\varepsilon'+(\bOmega\times\br_\perp)\cdot \bp']$.  This overall term changes the prefactor
$T=1/\beta$ to $1/\beta_\perp = \Gamma_\perp/\beta$.  The contribution from the second term involving $(\bOmega\times\br_\perp)\cdot\bp'$ is vanishing after the $\bp'$ integration.
%
Relabeling $\bp' \to \bp$, we finally get
\begin{equation}
  \calF = \frac{1}{\beta_\perp} \int \frac{d^3p}{(2\pi)^3} \, \ln(1 - e^{-\beta_\perp \varepsilon})\,.
  \label{eq:RKT_F}
\end{equation}

Interestingly, $\calF$ in the above form of Eq.~\eqref{eq:RKT_F} depends on $\Omega$ only through $\Gamma_\perp$ in $\beta_\perp = \beta / \Gamma_\perp$.  Thus, the angular momentum density reads:
\begin{equation}
  \calJ = -\frac{\partial \calF}{\partial \Omega} = -\frac{\partial \beta_\perp}{\partial \Omega} \cdot \frac{\partial \calF}{\partial \beta_\perp} \,.
  \label{eq:J_cl}
\end{equation}
We can easily calculate the first derivative in the right-hand side of Eq.~\eqref{eq:J_cl} as
\begin{equation}
  \frac{\partial \beta_\perp}{\partial \Omega} = -\frac{\beta}{\Gamma_\perp^2} \frac{\partial \Gamma_\perp}{\partial \Omega} = -\beta_\perp r_\perp^2 \Omega \Gamma^2_\perp\,.
  \label{eq:beta_deriv}
\end{equation}
For the second derivative in the right-hand side of Eq.~\eqref{eq:J_cl}, 
we can make use of the thermodynamic relation:
\begin{equation}
  \frac{\partial\mathcal{F}}{\partial\beta_\perp} = -\frac{1}{\beta_\perp^2} \frac{\partial \mathcal{F}}{\partial T_\perp}
  = \frac{s_\perp}{\beta_\perp^2} \,,
  \label{eq:F_deriv}
\end{equation}
where $s_\perp$ denotes the local entropy density.
We note that the entropy density, $s_\perp$, and the local enthalpy density, $h_\perp$, are related as $s_\perp / \beta_\perp = h_\perp$.  Putting together Eqs.~\eqref{eq:J_cl}, \eqref{eq:beta_deriv}, and \eqref{eq:F_deriv}, we arrive at
\begin{equation}
  \calJ = \Omega\, r_\perp^2 \Gamma_\perp^2 h_\perp \,.
\end{equation}
From this expression, we can immediately conclude our claim:
\begin{equation}
  \calI = \frac{\partial \calJ}{\partial\Omega} \biggr\rvert_{\Omega=0} =  r_\perp^2 h \,.
 \label{eq:rkt:I}
\end{equation}

For later convenience, let us write down the explicit forms of the thermodynamic quantities for a noninteracting Bose gas.  Using the function $h_n(y)$ introduced in Ref.~\cite{Kapusta:2007xjq}, whose definition is given in Eq.~\eqref{eq:hn_def} in Appendix~\ref{app:formulas}, we find the following expressions:
\begin{align}
 p &= \frac{1}{3} \int dP\, \bp^2 \nb(\varepsilon) = \frac{4T^4}{\pi^2} h_5(y) \,, \notag\\
 e &= \int dP\, \varepsilon^2 \nb(\varepsilon) = \frac{12 T^4}{\pi^2} h_5(y) + \frac{T^4 y^2}{\pi^2} h_3(y) \,, \notag\\
 h &= p + e = \frac{16 T^4}{\pi^2} h_5(y) + \frac{T^4 y^2}{\pi^2} h_3(y) \,,
 \label{eq:rkt:h}
\end{align}
where $\nb(\varepsilon)=1/(e^{\beta\varepsilon}-1)$ is the Bose-Einstein distribution function and $y= m / T$.  From these forms, using Eqs.~\eqref{eq:h35}, we can immediately derive the explicit forms in the high-temperature expansion under $T\gg m$ as
\begin{align}
  p &= \frac{\pi^2 T^4}{90} - \frac{T^2 m^2}{24} + \frac{T m^3}{12\pi}
  + \frac{m^4}{32\pi^2} \left(\gamma \! - \! \frac{3}{4} \!+\! \ln \frac{m}{4T} \right) + O(y^6) , \notag\\
  e &= \frac{\pi^2 T^4}{30} - \frac{T^2 m^2}{24}
  - \frac{m^4}{32\pi^2} \left(\gamma + \frac{1}{4} + \ln \frac{m}{4T}\right) + O(y^6)\,, \notag\\
  h &= \frac{2\pi^2 T^4}{45} -\frac{T^2 m^2}{12} + \frac{T m^3}{12\pi}
  - \frac{m^4}{32\pi^2} + O(y^6)\,.
 \label{eq:smallm}
\end{align}
Using these expressions, in the massless limit specifically, we have
\begin{equation}
 \mathcal{I}^{(0)}\rvert_{m = 0} = r_\perp^2 \, \frac{2\pi^2 T^4}{45} \,.
 \label{eq:I0}
\end{equation}

\subsection{Thermodynamic approach}
\label{sec:heuristic:thermo}

For the inclusion of interactions, it would be more convenient to employ a picture based on quantum field theory.  In the literature, a recipe to calculate thermodynamic quantities in a rotating frame has been established.  Here, we shall derive the moment of inertia for an interacting Bose gas using known expressions in finite-$T$ field theory.  A good starting point for the full thermodynamics is the 2PI expression of the free energy:
\begin{equation}
    \calF = \frac{1}{2} \sumint_{\:P} \bigl[ \ln G^{-1} + G_0^{-1} G - 1 \bigr] + \Phi \,,
    \label{eq:2PI}
\end{equation}
where $\sumint_{\,P}$ is a short-hand notation of finite-$T$ momentum-space integration, i.e.,
\begin{equation}
    \sumint_{\:P} = T\sum_{n=-\infty}^\infty \int \frac{d^3 p}{(2\pi)^3} \,.
    \label{eq:sumint}
\end{equation}
The last term in Eq.~\eqref{eq:2PI}, $\Phi$, represents the 2PI skeleton diagrams.  In the integrand, $G$ is the full propagator defined by $G^{-1} = G_0^{-1} + \Pi$ with the tree-level propagator, $G_0$, and the self-energy, $\Pi$.  The first term in Eq.~\eqref{eq:2PI} represents the resummed loop contribution, while the second term, $G_0^{-1} G$, is necessary to make the 2PI formalism self-consistent.

According to the recipe, if we consider thermodynamics in a rotating frame, we should modify the formulas in two places.  The first one lies in the modification in the phase space integral in the cylindrical coordinates, that is,
\begin{align}
  \frac{d^3 p}{(2\pi)^3}
  \to &\int_\ell \frac{d^3 p}{(2\pi)^3} \notag\\
  &= \frac{1}{4\pi^2} \! \int_0^\infty \!\! dp_\perp p_\perp \! \int_{-\infty}^{\infty} dp^z \!
  \sum_{\ell = -\infty}^\infty [J_\ell(p_\perp r_\perp)]^2 . \label{eq:dP_cylindrical}
\end{align}
Here, $J_n(x)$ represents the Bessel function of the first kind and $\ell$ corresponds to the angular momentum along $\bm{e}_z$, the rotation axis.  The second modification is found in the energy shift, i.e.,
\begin{equation}
    \varepsilon \;\to\; \modep = \varepsilon - \ell\Omega \,.
\end{equation}
This shift is more essential for rotation and this recipe brings the major dependence on $\Omega$ in thermodynamics; see Refs.~\cite{Jiang:2016wvv,Fujimoto:2021xix}.  In Eq.~\eqref{eq:2PI}, however, we should still take account of extra dependence on $\Omega$ through the self-energy, $\Pi$, for higher-order terms.

\subsubsection{Zeroth order contribution}
\label{sec:heuristic:lam0}

At zeroth order in $\lambda$, we drop the interaction effects, leading to $\Pi=\Phi=0$ and $G = G_0$.  Therefore, Eq.~\eqref{eq:2PI} reduces to $\calF \to \calF^{(0)} = \frac{1}{2} \sumint \ln G_0^{-1}$.  The above-mentioned recipe immediately gives the free energy of a noninteracting Bose gas under rotation as
\begin{equation}
    \calF^{(0)} = T \int_\ell\frac{d^3p}{(2\pi)^3} \, \ln\bigl( 1 - e^{-\beta \modep} \bigr) \,.
    \label{eq:F0}
\end{equation}
This expression can be evaluated up to $O(\Omega^2)$ by the following series expansion:
\begin{equation}
  \ln\bigl(1 - e^{-\beta \tilde{\varepsilon}}\bigr) = \left(1 - \Omega \ell \frac{d}{d\varepsilon} + \frac{\Omega^2 \ell^2}{2!} \frac{d^2}{d\varepsilon^2}\right) \ln\bigl(1 - e^{-\beta \varepsilon}\bigr) + O(\Omega^3) \,.
 \label{eq:log_series}
\end{equation}
From this expression, we obtain a formal form of the moment of inertia as
\begin{align}
    \calI^{(0)} = -\frac{\partial^2 \calF^{(0)}}{\partial\Omega^2} \biggr\rvert_{\Omega=0}
    = -\int_\ell \frac{d^3 p}{(2\pi)^3} \, \ell^2\, \frac{\partial n(\varepsilon)}{\partial\varepsilon}
    \label{eq:I0formal}\,.
\end{align}

Coming back to the series expansion in Eq.~\eqref{eq:log_series},
we can perform the sum over $\ell$ using the identities:
\begin{equation}
 \sum_{\ell = -\infty}^\infty J_\ell^2(z) = 1 \,, \quad
 \sum_{\ell = -\infty}^\infty J_\ell^2(z) \ell = 0\,, \quad
 \sum_{\ell = -\infty}^\infty J_\ell^2(z) \ell^2 = \frac{z^2}{2} \,,
 \label{eq:sum_ell}
\end{equation}
leading to
\begin{equation}
  \calF^{(0)} \simeq T \int \frac{d^3p}{(2\pi)^3} \left( 1 +
  \Omega^2 r_\perp^2
  \frac{p_\perp^2}{4} \frac{d^2}{d\varepsilon^2} \right) \ln\bigl(1 - e^{-\beta \varepsilon}\bigr) \,,
\end{equation}
where we neglected terms of order $O(\Omega^4)$.
Under the $d^3p$ integration, $p_\perp^2$ in the integrand can be replaced by $\frac{2}{3} \bp^2$ due to spherical symmetry.  Using the integration by parts, we arrive at
\begin{equation}
  \calF^{(0)} = -\int dP \left[\frac{\bm{p}^2}{3} + \frac{\Omega^2 r_\perp^2}{2} \left(\frac{\bm{p}^2}{3} + \varepsilon^2\right)\right] \nb(\varepsilon) \,.
  \label{eq:expanded_F0}
\end{equation}
It is easy to identify the pressure and enthalpy contributions in the above expression [see Eq.~\eqref{eq:rkt:h}] as
\begin{equation}
 \calF^{(0)} = -p - \frac{\Omega^2 r_\perp^2}{2} h + O(\Omega^4) \,,
\end{equation}
which immediately gives
\begin{equation}
  \calI^{(0)} = -\left.\frac{\partial^2 \calF^{(0)}}{\partial \Omega^2}\right\rvert_{\Omega = 0} = r_\perp^2 h \,.
\end{equation}

Let us make a remark about an explicit connection to the field-theoretical calculation.  To this end, it would be more instructive to step back to the formula before taking the Matsubara sum.  That is, the integrand in Eq.~\eqref{eq:F0} appears from
\begin{equation}
  \frac{1}{2}\sum_{n=-\infty}^\infty \ln \bigl[ (\omega_n - i\ell\Omega)^2 + \varepsilon^2 \bigr] \,,  
\end{equation}
where $\omega_n = 2\pi T n$ is the Matsubara frequency.  Assuming that the phase-space integration and the $\Omega$ derivative are commutable, we can find the moment of inertia as
\begin{equation}
  \calI^{(0)} = \frac{1}{2}\int_\ell \frac{dp^3}{(2\pi)^3} \Bigl( \mathcal{M}_{(1)} + \mathcal{M}_{(2)} \Bigr)
\end{equation}
with
\begin{equation}
    \mathcal{M}_{(1)} = -T\sum_n \biggl( \frac{-2i\ell \omega_n}{\omega_n^2 + \varepsilon^2} \biggr)^2 \,,\ \ 
    \mathcal{M}_{(2)} = -T\sum_n \frac{2\ell^2}{\omega_n^2 + \varepsilon^2} \,.
\end{equation}
These expressions correspond exactly to the leading order terms of $\langle \mathcal{L}_{(1)} \mathcal{L}_{(1)} \rangle$ and $\langle \mathcal{L}_{(2)} \rangle$, respectively, shown in Eq.~\eqref{eq:I1I2}.

Using Eqs.~\eqref{eq:sumn_D} and \eqref{eq:sumn_dD}, we can explicitly take the Matsubara sums to find
\begin{align}
    \mathcal{M}_{(1)} &= 2\ell^2 \frac{\partial \nb(\varepsilon)}{\partial \varepsilon} + \frac{\ell^2}{\varepsilon} \bigl[ 1 + 2\nb(\varepsilon) \bigr] \,,\\
    \mathcal{M}_{(2)} &= -\frac{\ell^2}{\varepsilon} \bigl[ 1 + 2\nb(\varepsilon) \bigr]\,,
\end{align}
which eventually leads to
\begin{equation}
    \calI^{(0)} = -\int_\ell \frac{d^3p}{(2\pi)^3}\, \ell^2 \frac{\partial n(\varepsilon)}{\partial \varepsilon} \,.
\end{equation}
This exactly recovers the formal expression in Eq.~\eqref{eq:I0formal}.

\subsubsection{Higher-order contribution}
\label{sec:heuristic:lam1}

In principle, we can systematically expand Eq.~\eqref{eq:2PI} to deduce the higher-order corrections.  Here, let us consider the corrections up to $O(\lambda^{3/2})$.  For this purpose, it is convenient to reorganize Eq.~\eqref{eq:2PI} as
\begin{align}
    \calF &= \frac{1}{2} \sumint_{\:P} \bigl[ \ln G_0^{-1} + ( \ln G^{-1} - \ln G_0^{-1}) - \Pi G] + \Phi\,.
    \label{eq:2PI2}
\end{align}
Here, we used $G_0^{-1}=G^{-1}-\Pi$ for the last term.  The first term is nothing but the zeroth order contribution of $\calF^{(0)}$ as discussed in the previous subsection.  The self-energy $\Pi$ originates from the loop diagrams, which admits an expansion in $\lambda$.  In the present treatment, we can drop diagrams of $O(\lambda^2)$ and higher orders.

Since detailed calculations follow in the next section, we shall simply refer to the known result of the self-energy correction (see, e.g., Ref.~\cite{Kapusta:2007xjq} for a standard textbook\footnote{Our convention of the interaction strength, $\lambda$, differs from $\lambda_\mathrm{KG}$ defined by Kapusta and Gale in Ref.~\cite{Kapusta:2007xjq} by a factor of $4!$, namely, $\lambda_\mathrm{KG} = \lambda/4!$.}).
Up to $O(\Omega^2, \lambda^{3/2})$, we have
\begin{equation}
    \Pi(\Omega) \simeq \left[ \frac{\lambda T^2}{4!} - \frac{3 T^2}{\pi}\left(\frac{\lambda}{4!}\right)^{3/2} \right] ( 1 + r_\perp^2 \Omega^2 )\,.
    \label{eq:selfenergy_rot}
\end{equation}
Here, for simplicity, we show the expression only for the massless theory.
Without the rotation-induced terms, $r_\perp^2 \Omega^2$, this is a standard expression of the self-energy from the one-loop integral in the 2PI formalism in which the in-medium mass correction, $m^2 \to m_\Pi^2 = m^2 + \Pi$, is included in the dressed propagator; see Eq.~\eqref{eq:Pi_small} for the derivation.  It is nontrivial that the rotation induced terms can be factorized in such a way as in Eq.~\eqref{eq:selfenergy_rot}, and the above form for the massless case is derived by explicit calculations in Appendix~\ref{App:Pi_Omega}.

The term $\mathcal{F}_{\Pi G} = -\frac{1}{2} \sumint \Pi G$ in Eq.~\eqref{eq:2PI2} is the easiest to estimate. The leading-order self-energy has no momentum dependence, so this last term is factorized as a product of $\Pi$ and the loop integral of $G$, that is nothing but the one-loop self-energy: $\Pi = \frac{\lambda}{2} \sumint G$.  In this way, up to $O(\Omega^2, \lambda^{3/2})$, this last term takes the form of
\begin{equation}
    \calF_{\Pi G} = -\frac{1}{\lambda} \Pi^2 \simeq \left[-\frac{\lambda}{4!} \frac{T^4}{24} + \left(\frac{\lambda}{4!}\right)^{3/2} \frac{T^4}{4\pi}\right] (1 + 2 r_\perp^2 \Omega^2)\,.
\end{equation}

The second term in Eq.~\eqref{eq:2PI2} is the most nontrivial correction of our interest. 
It is possible to express this term as the difference between two noninteracting free energies, i.e.,
\begin{equation}
  \calF_\mathrm{log} = \frac{1}{2} \sumint_{\:P} (\ln G^{-1} - \ln G_0^{-1}) = \calF^{(0)}(m_\Pi) - \calF^{(0)}(m) \,,
\end{equation}
where $m_\Pi^2 = m^2 + \Pi$ is the dressed mass.
At vanishing mass $m \to 0$, we can obtain perturbative expressions for $\calF_\mathrm{log}$ using Eq.~\eqref{eq:smallm} and the results are, up to $O(\Omega^2, \lambda^{3/2})$,
\begin{equation}
    \calF_\mathrm{log} \simeq \biggl[ \frac{\lambda}{4!} \frac{T^4}{24} - \left(\frac{\lambda}{4!}\right)^{3/2} \frac{5T^4}{24\pi} \biggr] (1 + 2r_\perp^2 \Omega^2 ) \,.
\end{equation}
We see that the terms of $O(\lambda)$ cancel exactly once $\calF_{\Pi G}$ and $\calF_\mathrm{log}$ are added together.  This cancellation is necessary to avoid double counting in the 2PI formalism, and the genuine corrections of $O(\lambda)$ should come from the leading contribution in $\Phi$.

We can obtain the $\Phi$ contributions up to $O(\lambda^{3/2})$ from the nonperturbative relation $\delta \Phi/\delta G=\frac{1}{2}\Pi$, from which we see that
\begin{equation}
  \Phi = -\frac{1}{2} \calF_{\Pi G}
\end{equation}
in the leading order where $\Pi$ has no momentum dependence.

Putting all the results together, we finally arrive at the following expression in the massless case:
\begin{equation}
 \calF \simeq \left[-\frac{\pi^2 T^4}{90} + \frac{\lambda}{4!} \frac{T^4}{48} - \left(\frac{\lambda}{4!}\right)^{3/2} \frac{T^4}{12\pi}\right](1 + 2r_\perp^2 \Omega^2)
 \label{eq:F_heuristic}
\end{equation}
up to $O(\lambda^{3/2},\Omega^2)$.  Accordingly, the moment of inertia can be expanded perturbatively as
\begin{equation}
 \calI = \calI^{(0)} + \calI^{(1)} + \calI^{(3/2)} + O(\lambda^2)
 \label{eq:Iexp}
\end{equation}
with $\calI^{(0)}$ given in Eq.~\eqref{eq:I0}, while the corrections due to interactions read:
\begin{equation}
    \calI^{(1)} = -r_\perp^2 \, \frac{\lambda}{4!} \frac{T^4}{12} \,,
    \qquad
    \calI^{(3/2)} = r_\perp^2 \, \left(\frac{\lambda}{4!}\right)^{3/2} \frac{T^4}{3\pi} \,.
    \label{eq:Icorr}
\end{equation}
In the massless scalar theory, the loop corrections to the pressure are known, from which the enthalpy density can be obtained as
\begin{equation}
    h = 4p = \frac{2\pi^2 T^4}{45} - \frac{\lambda}{4!}\frac{T^4}{12} + \left(\frac{\lambda}{4!}\right)^{3/2} \frac{T^4}{3\pi} + O(\lambda^2) \,.
    \label{eq:h_heuristic}
\end{equation}
Comparing Eqs.~\eqref{eq:I0}, \eqref{eq:Iexp}, \eqref{eq:Icorr}, and \eqref{eq:h_heuristic}, we can verify our claim, $\calI = r_\perp^2 h$, in an explicit manner. 

Although the thermodynamic approach is intuitive, it is unavoidable to treat the higher-loop contributions from $\Phi$ in order to consider corrections beyond $O(\lambda^{3/2})$.  Then, it would be more systematic to take the expression in terms of the correlation functions as a starting point and develop a fully diagramatic method.
This clearly motivates our discussion in the next section.

\section{Loop computation}
\label{sec:loop}

We formulate the computation of the moment of inertia using Eqs.~\eqref{eq:I1} and \eqref{eq:I2} with the correlation functions, $G_2$ and $G_4$, directly.  Although the correlation functions are ordinary building blocks in quantum field theory, the moment of inertia needs nonlocal integrations, which requires a careful treatment in coordinate space. 

In Subsec~\ref{sec:loop:D2}, we review the two-point function with the self-energy correction under the 2PI formalism \cite{Fejos:2011iy, Andersen:2002jz}. The enthalpy in the self-interacting theory is derived in Subsec.~\ref{sec:loop:h}. The $I_2$ and $I_1$ contributions to the moment of inertia are evaluated in Subsecs.~\ref{sec:loop:I1} and \ref{sec:loop:I2}, respectively. The total moment of inertia is computed in Subsec.~\ref{sec:loop:I}, where we also present a comparison with the results of Ref.~\cite{Siri:2024scq}.

\subsection{Two-point function and subtraction scheme}
\label{sec:loop:D2}

We will consider the renormalized perturbation theory in a standard way as
\begin{equation}
 \mathcal{L} = \frac{1}{2} \partial_\mu \phi \partial^\mu \phi - \frac{1}{2} m^2 \phi^2 - \frac{\lambda}{4!} \phi^4 - \frac{1}{2} \delta_m \phi^2 \,,
 \label{eq:Lren}
\end{equation}
where the counter terms associated with the wavefunction renormalization, $\delta Z$, and the coupling renormalization, $\delta_\lambda$, are neglected because they appear only from $O(\lambda^2)$ corrections.

We consider the Fourier transformed two-point function defined by
\begin{equation}
  \propx(X - Y) 
  = \sumint_{\:P} e^{i P \cdot (X - Y)}\, \propp(P) \,,
\end{equation}
where we use the short-hand notation introduced in Eq.~\eqref{eq:sumint}.  We note that $P$ is the Euclidean four-momentum, $P_\mu = (\omega_n, \bp)$.  The free propagator reads:
\begin{equation}
  \propp_0(P) = 
\begin{tikzpicture}[baseline={([yshift=-.5ex]x)}]
  \begin{feynman}[small]
    \vertex [] (x) {};
    \vertex [right=of x] (y) {};
    \diagram* {
      (x) -- [fermion, edge label'=$P$] (y),
    };
  \end{feynman}
\end{tikzpicture}
 = \frac{1}{\omega_n^2 + \varepsilon^2}
 = \frac{1}{\omega_n^2 + \bp^2 + m^2} \,.
\end{equation}
By definition of self-energy, the full propagator (apart from the wavefunction renormalization) is given as
\begin{equation}
    \propp(P) = 
\begin{tikzpicture}[baseline={([yshift=-.5ex]x)}]
  \begin{feynman}[small]
    \vertex [] (x) {};
    \vertex [right=of x] (y) {};
    \diagram* {
      (x) -- [double, with arrow=0.5, edge label'=$P$] (y),
    };
  \end{feynman}
\end{tikzpicture}
    = \frac{1}{\omega_n^2 + \bp^2 + m^2 + \selfp(P)}\,.
\end{equation}
Thus, for the propagator in the resummed perturbation theory, we need to evaluate the self-energy $\selfp(P)$, which in general depends on the external momentum $P$.

The leading-order contribution to the self-energy appears from the one-loop diagram with the dressed propagator:
\begin{align}
  \Pi^{(1)} &= - \begin{tikzpicture}[baseline={([yshift=-.5ex]z)}]
    \begin{feynman}[small]
      \vertex [] (x) {};
      \vertex [right=of x, dot] (z) {};
      \vertex [right=of z] (y) {};
      \diagram* {
        (x) -- [plain] (z) -- [plain] (y),
      };
      \draw [double] (z) arc [start angle=-90, end angle=270, radius=0.3cm];
    \end{feynman}
  \end{tikzpicture} - \begin{tikzpicture}[baseline={([yshift=-.5ex]z)}]
    \begin{feynman}[small]
      \vertex [] (x) {};
      \vertex [right=of x, crossed dot] (z) {};
      \vertex [right=of z] (y) {};
      \diagram* {
        (x) -- [plain] (z) -- [plain] (y),
      };      
    \end{feynman}
  \end{tikzpicture} \nonumber \\
  &= \frac{\lambda}{2} \propx(0) + \delta_m \,,
  \label{eq:Pi_tadpole}
\end{align}
where the counterterm $\delta_m$ subtracts the vacuum contribution~\cite{Kapusta:2007xjq, Laine:2016hma}.
Since $G(0)$ contains $\Pi$, the above self-energy expression can be understood as the self-consistent condition (see, e.g., Eqs.~(2.27) and (2.29) in Ref.~\cite{Blaizot:2000fc}), that is,
\begin{equation}
  \Pi = \frac{\lambda}{2} \int dP_{\Pi}\, \nb(\varepsilon_\Pi) \,, \qquad 
  \int dP_{\Pi} = \int \frac{d^3p}{(2\pi)^3 \varepsilon_\Pi}\,,
 \label{eq:Pi}
\end{equation}
where $\varepsilon_\Pi^2 = \bp^2 + m_\Pi^2$ and $m_\Pi^2 = m^2 + \Pi$.
In the perturbation theory, we can solve this self-consistent condition by the iterative procedures.  For $\Pi$ in the right-hand side, we use $\Pi\simeq \Pi^{(1)}$, and then the momentum integration multiplied by $\lambda/2$ gives the next-order correction corresponding to the ring diagrams; see Appendix~\ref{app:Pi}.
In this way, we can find the perturbative expansion up to $O(\lambda^{3/2})$ correction as
\begin{equation}
  \Pi_\mathrm{pert} = \frac{\lambda T^2}{4!} - \frac{3T^2}{\pi} \left(\frac{\lambda}{4!}\right)^{3/2} + O(\lambda^2) \,,
 \label{eq:Pi_small}
\end{equation}
whose second term coincides with that of the resummed self-energy $\Pi^{(3/2)}$, shown in Eq.~\eqref{eq:Pi32_small}.
This form of $\Pi_\mathrm{pert}$ defines the perturbatively resummed propagator, $\widetilde{G}_\mathrm{pert}(P)$.

In principle, we can repeat this iterative procedure again.  Alternatively we can treat Eq.~\eqref{eq:Pi} as a gap equation and numerically solve it to obtain $\Pi_{\text{resum}}$.  Although we do not consider diagrammatic contribution of $O(\lambda^2)$ in the 2PI expansion, $\Pi_{\text{resum}}$ contains a special class of higher-order diagrams called daisy diagrams that make the exact result in the large-$N$ limit in the $\mathrm{O}(N)$ scalar theory~\cite{Amelino-Camelia:1997xip}.

We note that the resummed self-energy, $\Pi_\mathrm{resum}$, is independent of external momenta, 
and the momentum dependence appears first from the sunset diagram at $O(\lambda^2)$.  After all, using the momentum-independent self-energy, i.e., either $\Pi_\mathrm{pert}$ or $\Pi_\mathrm{resum}$, we can approximate the nonperturbative propagator as
%
%
\begin{equation}
 \widetilde{G}_\mathrm{pert/resum}(P) = \frac{1}{\omega_n^2 + \bp^2 + m^2 + \Pi_\mathrm{pert/resum}} \,.
 \label{eq:Dpi}
\end{equation}

\subsection{Enthalpy computation}\label{sec:loop:h}

The enthalpy density $h = p + e$ of a Bose gas can be evaluated starting from the energy-momentum tensor given in Eq.~\eqref{eq:M_Theta}.
We emphasize that this is a technically important point; according to the 2PI formalism in Eq.~\eqref{eq:2PI}, the simple generalization of the noninteracting one-loop calculation with the full propagator, i.e., $\mathrm{tr}\ln G_0^{-1}\to \mathrm{tr}\ln G^{-1}$, cannot reproduce even the lowest-order perturbation theory as argued in Refs.~\cite{Baier:1999db,Andersen:2002jz,Fujimoto:2020tjc}.  As we see below, the energy-momentum tensor is quite convenient for the consistent resummation; see also Ref.~\cite{Blaizot:2000fc} for a complementary approach within the 2PI method.

Identifying the energy density and the pressure via
\begin{align}
  e &= \langle \hat{\Theta}^{tt} \rangle = e' - \langle \hat{\mathcal{L}} \rangle \,, \notag\\
  p &= \frac{1}{3}\langle\hat{\Theta}^{xx} + \hat{\Theta}^{yy} + \hat{\Theta}^{zz}\rangle = p' + \langle \hat{\mathcal{L}} \rangle
\end{align}
with $e' = \langle (\partial_t \hat{\phi})^2\rangle$ and $p' = \langle \tfrac{1}{3} (\bnabla \hat{\phi})^2 \rangle$.
it is easy to see that 
\begin{equation}
  h = e + p = e' + p' = \langle (\partial_t \phi)^2 + \tfrac{1}{3} (\bnabla \phi)^2 \rangle.
\end{equation}
Splitting the operator product by introducing the auxiliary point $X'$, we can write
\begin{align}
  h(X)
  &= \int_{X'} \delta^{(4)}(X - X') \left(-\partial_\tau \partial_\tau' + \tfrac{1}{3} \bnabla \cdot \bnabla'\right) G(X - X') \notag\\
  &= \sumint_{\:P} (-\omega_n^2 + \tfrac{1}{3} \bm{p}^2) \widetilde{G}(P) \,,
 \label{eq:hgen}
\end{align}
where we changed $\partial_t\to i\partial_\tau$.

Substituting $\widetilde{G}(P)$ from Eq.~\eqref{eq:Dpi} and using Eqs.~\eqref{eq:sumn_D} and \eqref{eq:sumn_omegan}, we obtain 
\begin{subequations}
\begin{align}
  e' &= \int dP_\Pi\, \varepsilon_\Pi^2 \, \nb(\varepsilon_\Pi) \,,
  \label{eq:epspi}\\
  p' &= \int dP_\Pi\, \frac{\bp^2}{3} \,  \nb(\varepsilon_\Pi) \,,
 \label{eq:Ppi}
\end{align}
\end{subequations}
where we dropped the vacuum terms with two ultraviolet divergences.  One is the contact term as pointed out in Appendix~\ref{app:formulas}, and the other is the $T=0$ contribution that is not considered in this work.

To evaluate the full pressure $p$, we now evaluate the expectation value of the Lagrangian \eqref{eq:L} as follows:
\begin{equation}
  \langle \hat{\mathcal{L}} \rangle \simeq -\frac{1}{2} \sumint_{\:P} (\omega_n^2 + \bp^2 + m^2) \widetilde{G}(P) - \frac{\lambda}{8} \langle \hat{\phi}^2 \rangle^2 \,,
  \label{eq:Lavg_aux}
\end{equation}
where we approximated $\langle \hat{\phi}^4 \rangle \simeq 3 \langle \hat{\phi}^2 \rangle^2 + O(\lambda)$, which is consistent with the perturbative expansion up to $O(\lambda^{3/2})$, or the daisy diagram resummation, that we employed throughout this work.  We note that $\Pi = \frac{\lambda}{2} \langle \hat{\phi}^2 \rangle$ simplifies the latter term.
Performing the Matsubara sum in Eq.~\eqref{eq:Lavg_aux} using Eqs.~\eqref{eq:sumn_omegan} and \eqref{eq:sumn_D}, we arrive at
\begin{equation}
  \langle \hat{\mathcal{L}} \rangle \simeq \frac{1}{2} \int dP_\Pi \, (\varepsilon_\Pi^2 - \bp^2 -m^2) \nb(\varepsilon_\Pi) - \Pi^2 \frac{1}{2\lambda} \,.
\end{equation}
Using $\varepsilon_\Pi^2 - \bp^2 -m^2 = \Pi$ and Eq.~\eqref{eq:Pi}, 
we finally find the Lagrangian expectation value as
\begin{equation}
 \langle \hat{\mathcal{L}} \rangle = \frac{1}{2\lambda} \Pi^2 \,.
\end{equation}
We therefore obtain
\begin{equation}
  e = 3p' + \frac{2m^2}{\lambda} \Pi + \frac{3}{2\lambda} \Pi^2 \,,
  \qquad 
  p = p' + \frac{1}{2\lambda} \Pi^2 \,.
 \label{eq:eps_P_full}
\end{equation}
In the $m\to 0$ limit, the full pressure can be expanded as
\begin{equation}
  p = \frac{\pi^2 T^4}{90} - \frac{\lambda}{4!} \frac{T^4}{48} + \left(\frac{\lambda}{4!}\right)^{3/2} \frac{T^4}{12\pi} + O(\lambda^2) \,,
\end{equation}
which agrees with $-\calF|_{\Omega=0}$ in Eq.~\eqref{eq:F_heuristic}.  It is also easy to confirm $e = 3p$ and $h = 4p$ at $m=0$, which is in agreement with the result in Eq.~\eqref{eq:h_heuristic}.

\subsection{\texorpdfstring{$I_2$}{I2} contribution}\label{sec:loop:I1}

Let us consider the diagrammatic evaluation of the moment of inertia.  We begin with $I_2$, for it is technically straightforward.

Substituting the Fourier transformed expression for $G(X - X')$ and performing the $X'$ integration in Eq.~\eqref{eq:I2}, we find
\begin{equation}
  I_2 = T \int_X \sumint_{\:P} (y p^x - x p^y)^2 \widetilde{G}(P) \,.
\end{equation}
Under the $X$ integration, due to cylindrical symmetry in the $x$-$y$ plane, we can replace $x^2, y^2 \to r_\perp^2 / 2$ and $xy \to 0$.  Since $\widetilde{G}(P)$ has no special direction and it is a function of $P^2$, we can also replace $p_x^2, p_y^2 \to \frac{1}{3} \bm{p}^2$, such that 
\begin{equation}
  \calI_2 = \frac{dI_2}{d^3X} = r_\perp^2 \sumint_{\:P} \frac{\bp^2}{3} \widetilde{G}(P) \,,
\end{equation}
which coincides with the second term in Eq.~\eqref{eq:hgen}. Using $\widetilde{G}(P)$ in Eq.~\eqref{eq:Dpi} and the summation formula \eqref{eq:sumn_D} gives 
\begin{equation}
  \calI_2 = r_\perp^2 \, p'
 \label{eq:I2_res}
\end{equation}
with $p'$ given in Eq.~\eqref{eq:Ppi}.

\subsection{\texorpdfstring{$I_1$}{I1} contribution}\label{sec:loop:I2}

In order to compute $I_1$ in Eq.~\eqref{eq:I1}, we perform the Fourier transformation of the $n$-point function using
\begin{equation}
  \begin{split}
  &G(X_1, \dots X_n) \\
  &\qquad = \sumint_{\:P_1} \cdots \sumint_{\:P_n} e^{i\sum_{i= 1}^n P_i \cdot X_i} 
 \widetilde{G}(P_1, \dots P_n) \,.
   \end{split}
\end{equation}
Using the above expression for $n=4$ in Eq.~\eqref{eq:I1} gives
\begin{equation}
  \begin{split}
  I_1 &= -T \int_{X_1, X_2} \sumint_{\:P_1} \sumint_{\:P_2} \sumint_{\:P_{1'}} \sumint_{\:P_{2'}} \\
  &\quad \times 
  \omega_1 \omega_2 (-y_1 p^x_{1'} + x_1 p^y_{1'})(-y_2 p^x_{2'} + x_2 p^y_{2'}) \\
  &\quad \times 
  e^{i(P_1 + P_{1'}) \cdot X_1 + i(P_2 + P_{2'}) \cdot X_2} \widetilde{G}(P_1, P_{1'}, P_2, P_{2'}) \,.
 \label{eq:I1_D4_P}
  \end{split}
\end{equation}

In general, the four-point function has contributions from the disconnected and connected diagrams, denoted by $G_\mathrm{disc}$ and $G_\mathrm{conn}$, respectively, that is,
\begin{equation}
 G(X_1, X_{1'}, X_2, X_{2'}) = G_\mathrm{disc} + G_\mathrm{conn} \,.
\end{equation}
The connected diagram contribution appears from the interaction and the first-order term is
\begin{equation}
 G_\mathrm{conn}^{(1)} = \begin{tikzpicture}[scale=0.8, transform shape, baseline={([yshift=-.5ex]z)}]
    \begin{feynman}[small]
    \vertex [dot] (z) {};
    \vertex [above left=of z] (x1);
    \vertex [above right=of z] (x2);
    \vertex [below left=of z] (x3);
    \vertex [below right=of z] (x4);
    
    \diagram* {
      (x1) -- [double] (z),
      (x2) -- [double] (z),
      (x3) -- [double] (z),
      (x4) -- [double] (z),
    };
     \end{feynman}
\end{tikzpicture} = -\lambda \int d^4Z \prod_{i = 1,1',2,2'} G(X_i - Z) \,.
  \label{eq:Gcon1}
\end{equation}
It should be noted that any diagrams including additional vertices would make contributions of $O(\lambda^2)$ or higher orders, which are not considered in this work.
Now, we are showing that the moment of inertia from $G_\mathrm{conn}^{(1)}$ is vanishing, i.e., $\calI_{1;\mathrm{conn}}^{(1)}=0$ below.

From Eq.~\eqref{eq:Gcon1}, we can write down the Fourier transformed expression as
\begin{equation}
  \widetilde{G}^{(1)}_\mathrm{conn} = -\lambda \beta(2\pi)^3 \delta^{(4)}(P_1 + P_2 + P_{1'} + P_{2'})\, \widetilde{G}_1\, \widetilde{G}_2\, \widetilde{G}_{1'}\, \widetilde{G}_{2'} \,,
\end{equation}
where $\widetilde{G}_i = \widetilde{G}(P_i)$ with $i=1,1',2,2'$.
Substituting this back into Eq.~\eqref{eq:I1_D4_P} and performing the $X_2$ integration lead to
\begin{equation}
  I_{1;\mathrm{conn}}^{(1)} = \frac{\lambda}{2} \int d^3x\, r_\perp^2 \sumint_{\:P_1} \sumint_{\:P_2} \omega_1 \omega_2 \bp^\perp_1 \cdot \bp^\perp_2\, [\widetilde{G}_1 \widetilde{G}_2]^2
\label{eq:I1_van}
\end{equation}
with $\bp^\perp_1 \cdot \bp^\perp_2 = p^x_1 p^x_2 + p^y_1 p^y_2$. The integrand is an odd function of $P^\mu_{i}$, thus $\sumint_{\:P_i}$ results in zero.  Hence, we conclude $I_{1;\mathrm{conn}}^{(1)} = 0$. 

Up to the order of our present interest, therefore, we are left with the contributions coming from the disconnected diagrams.
In the leading order, the disconnected contribution is given by all possible combinations of two-point functions that connect the four points.  Thus, $G_\mathrm{disc} = G_{12} G_{1'2'} + G_{11'} G_{22'} + G_{12'} G_{21'}$
with $G_{ij} = G(X_i - X_j)$ being the fully-dressed two-point function.  The Fourier transformed form is
\begin{equation}
  \begin{split}
  \widetilde{G}_\mathrm{disc} = \beta^2 & (2\pi)^6 [ \widetilde{G}_1 \widetilde{G}_2 \delta^{(4)}(P_1 + P_{1'}) \delta^{(4)}(P_2 + P_{2'}) \\
  & + \widetilde{G}_1 \widetilde{G}_{1'} \delta^{(4)}(P_1 + P_2) \delta^{(4)}(P_{1'} + P_{2'})\\
  & + \widetilde{G}_1 \widetilde{G}_2 \delta^{(4)}(P_1 + P_{2'}) \delta^{(4)}(P_2 + P_{1'})] \,.
   \end{split}
\end{equation}
These three terms correspond to distinct diagrams connecting (1) $1$-$1'$ and $2$-$2'$; (2) $1$-$2$ and $1'$-$2'$; and (3) $1$-$2'$ and $2$-$1'$.
We refer to these contributions, respectively, by $I_{1;1'}$, $I_{1;2}$, and $I_{1;2'}$ in what follows below.  The first one actually vanishes,
\begin{multline}
  I_{1;1'} = -T \int_{X_1, X_2} \sumint_{\:P_1} \sumint_{\:P_2} \omega_1 \omega_2 \\
  \times (y_1 p^x_1 - x_1 p^y_1)(y_2 p^x_2 - x_2 p^y_2)\widetilde{G}_1 \widetilde{G}_2 = 0 \,,
\end{multline}
as the integrand is odd with respect to all the momenta, as well as the transverse coordinates.  The other two terms make equal contributions;
$I_{1;2}=I_{1;2'}$.  In the case of the computation of $I_{1;2'}$, we have the following expression:
\begin{multline}
  I_{1;2'} = -T \int_{X_1, X_2} \sumint_{\:P_1} \sumint_{\:P_2} \omega_1 \omega_2 \widetilde{G}_1 \widetilde{G}_2 \\
  \:\times
  (-y_1 p^x_2 + x_1 p^y_2)(-y_2 p^x_1 + x_2 p^y_1)\, e^{i(P_1 - P_2) \cdot (X_1 - X_2)} \,.
\end{multline}
This integral does not vanish because of the remaining exponential factor.
We can perform the $X_2$ integration as 
\begin{multline}
  \int_{X_2} (-y_2 p^x_1 + x_2 p^y_1) e^{-i(P_1 - P_2) \cdot X_2} \\ 
 = \beta (2\pi)^3
 \left(-i \frac{\partial}{\partial p^y_1} p^x_1 + i \frac{\partial}{\partial p^x_1} p^y_1\right) \delta^{(4)}(P_1 - P_2) \,.
\end{multline}
Since the derivative of Dirac's delta function is defined by the integration by parts, the derivatives with respect to $p^x_1$ and $p^y_1$ should act on the remaining $P^\mu_{1}$ dependence in the integrand, namely on $\widetilde{G}_1 e^{i P_1 \cdot X_1}$. Due to the term $-y_1 p^x_2 + x_1 p^y_2$ in the integrand, the $X_1$ integration vanishes unless the $P^\mu_{1}$ derivatives act on the exponential factor, $e^{i P_1 \cdot X_1}$.
Hence, after performing the $X_2$ and $P_2$ integrations and renaming $X_1\to X$ and $P_1\to P$, we can simplify the integral as
\begin{align}
  I_{1;2'} &= -T \int_X \sumint_{\:P} \omega_n^2 \, [\widetilde{G}(P)]^2 (-y p^x + x p^y)^2 \nonumber\\
 &= -\frac{1}{3} \int d^3x\, r_\perp^2 \sumint_{\:P} \omega_n^2 \, \bp^2 [\widetilde{G}(P)]^2 \,,
\end{align}
where we replaced $x^2, y^2 \to \frac{1}{2} r_\perp^2$, $xy \to 0$, as well as $p_x^2 + p_y^2 \to \frac{2}{3} \bp^2$. 

To perform the Matsubara sum,
we use the explicit form of $\widetilde{G}_\mathrm{pert}(P)$ or $\widetilde{G}_\mathrm{resum}(P)$ in Eq.~\eqref{eq:Dpi}.
Switching from cylindrical to spherical coordinates in momentum space, we write $p [\widetilde{G}(P)]^2 = -\frac{1}{2} \partial_p \widetilde{G}(P)$. Integrating by parts with respect to $p=|\bp|$, we get
\begin{equation}
  I_{1;2'} = -\frac{1}{2} \int d^3x\, r_\perp^2 \sumint_{\:P} \omega_n^2 \widetilde{G}(P)  
  = \frac{1}{2} \int d^3x\, r_\perp^2\, e'
\end{equation}
with $e'$ introduced in Eq.~\eqref{eq:epspi}.  A similar calculation reveals $I_{1;2} = I_{1;2'}$, such that $\mathcal{I}_1 = d I_1 / d^3x$ becomes 
\begin{equation}
 \calI_1 = r_\perp^2\, e'\,.
 \label{eq:I1_res}
\end{equation}

\begin{figure}
    \centering
    \begin{tabular}{c}
    \includegraphics[width=\linewidth]{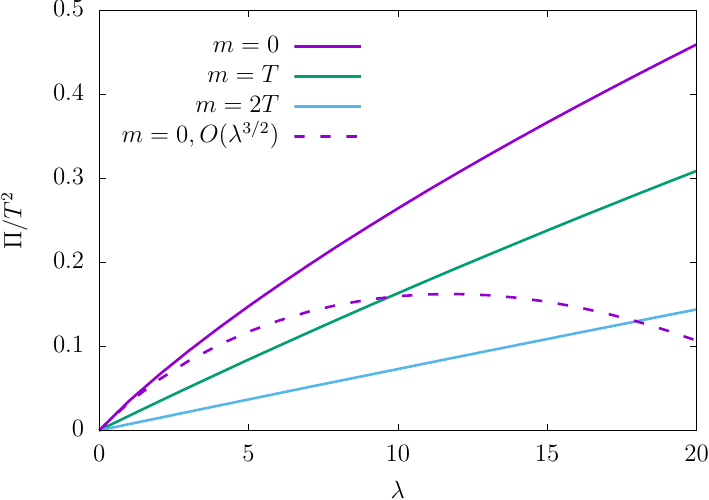} \\
    \includegraphics[width=\linewidth]{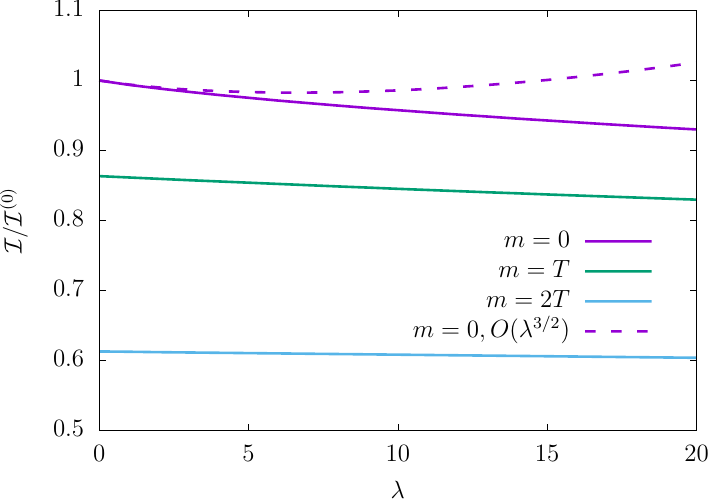} 
    \end{tabular}
    \caption{
    (Top) The self-energy $\Pi$ in Eq.~\eqref{eq:Pi} made dimensionless with $T^2$; (bottom) the moment of inertia $\calI$ given in Eq.~\eqref{eq:I_final} normalized with respect to $\calI^{(0)}$ in Eq.~\eqref{eq:I0} for the massless noninteracting Bose gas.  Three solid curves represent the results for different masses, while the dashed curves represent the perturbatively expanded results up to $O(\lambda^{3/2})$, as given in Eqs.~\eqref{eq:Pi_small} and \eqref{eq:I_final_smallm}.}
    \label{fig:momI20}
\end{figure}

Adding together $\calI_1$ in Eq.~\eqref{eq:I1_res} and $\calI_2$ in Eq.~\eqref{eq:I2_res} shows that, up to the terms that correctly produce contributions of $O(\lambda^{3/2})$, the moment of inertia density for a self-interacting scalar field reads:
\begin{equation}
 \calI = r_\perp^2 h\,, \qquad 
 h = \frac{16 T^4}{\pi^2} h_5(y_\Pi) + \frac{T^4 y_\Pi^2}{\pi^2} h_3(y_\Pi)
 \label{eq:I_final}
\end{equation}
with $y_\Pi = \sqrt{m^2 + \Pi} / T$.  The above relation holds for any value of the mass $m$ of the field quanta.  If we use $G_\mathrm{pert}(P)$ with $\Pi_\mathrm{pert}$, we can expand the final expression 
up to $O(\lambda^{3/2})$ and reproduce $\calI$ in the massless limit as
\begin{equation}
  \calI\rvert_{m = 0} \simeq r_\perp^2 \left[\frac{2\pi^2 T^4}{45} - \frac{\lambda}{4!}\frac{T^4}{12} + \left(\frac{\lambda}{4!}\right)^{3/2} \frac{T^4}{3\pi}\right]\,.
 \label{eq:I_final_smallm}
\end{equation}
Interestingly, if we use $G_\mathrm{resum}(P)$ with $\Pi_\mathrm{resum}$ numerically solved from Eq.~\eqref{eq:Pi}, we can compute the resummed result for $\calI$, as shown in Fig.~\ref{fig:momI20}.

The top and bottom panels of Fig.~\ref{fig:momI20} display the self-energy and the moment of inertia, respectively, for various values of the vacuum mass $m$.  The solid and dashed curves represent the results with $G_\mathrm{resum}(P)$ and $G_\mathrm{pert}(P)$, which deviate from each other for $\lambda \gtrsim 1$.
The self-energy is made dimensionless with $T^2$, while $\calI$ is normalized with respect to $\calI^{(0)}$ of the massless and noninteracting Bose gas given in Eq.~\eqref{eq:I0}.  Although the perturbative results expanded up to $O(\lambda^{3/2})$ in Eqs.~\eqref{eq:Pi_small} and \eqref{eq:I_final_smallm} exhibit nonmonotonic behavior, signaling the breakdown of na\"{i}ve perturbation, the resummed results with the self-consistent solution of the gap equation~\eqref{eq:Pi} show stable and monotonic behavior with modest dependence on the coupling strength $\lambda$. 
In the first-order 2PI formalism, at large coupling $\lambda \to \infty$, $\Pi \propto (\ln \lambda)^2$ grows and $h \propto (\ln \lambda)^4/ \lambda$ decreases, as discussed in Appendix~\ref{app:largelambda}.

\subsection{Total moment of inertia} \label{sec:loop:I}

The total moment of inertia inside a cylinder of height $H$ and radius $R$ can be obtained as
\begin{equation}
 I(R) = \int d^3x\, \mathcal{I}(\bm{r}_\perp)
 = \frac{\pi}{2} R^4 H\, h = \frac{V R^2}{2} h \,,
 \label{eq:IR}
\end{equation}
where $V=\pi R^2 H$.
This expression can be found equivalently by first integrating the free energy density $\calF$ over the volume of the cylinder and then differentiating twice with respect to $\Omega$.

We illustrate the above discussion for the computation of the total free energy $F^{(0)} = \int d^3x\, \calF^{(0)}$ of the noninteracting gas, starting from Eq.~\eqref{eq:F0}:
\begin{equation}
  \begin{split}
 F^{(0)} &\simeq \frac{T}{4\pi^2} \int d^3x \int_0^\infty dp_\perp p_\perp \int_{-\infty}^\infty dp^z \sum_{\ell = -\infty}^\infty \\
 &\quad\times [J_\ell(p_\perp r_\perp)]^2 
 \left( 1 + \frac{\Omega^2}{2} \ell^2 \frac{d^2}{d\varepsilon^2} \right) \ln(1 - e^{-\beta \varepsilon})\,.
  \end{split}
\end{equation}
Using the orthogonality relation,
\begin{equation}
 \int_0^\infty dr_\perp\, r_\perp J_\ell(p_\perp r_\perp) J_\ell(p_\perp' r_\perp) = \frac{\delta(p_\perp - p_\perp')}{p_\perp}\,,
 \label{eq:ortho}
\end{equation}
and recalling the summation formulas in Eq.~\eqref{eq:sum_ell}, we find the following expressions:
\begin{subequations}
\begin{align}
 \int d^3x \sum_{\ell = -\infty}^\infty J^2_\ell(p_\perp r_\perp) &= V\,, \\
 \int d^3x \sum_{\ell = -\infty}^\infty \ell^2 J^2_\ell(p_\perp r_\perp) &= \frac{p_\perp^2 R^2}{4} V\,. 
\end{align}
\label{eq:sum_ell_int}
\end{subequations}
Therefore, we can expand the spatially integrated free energy in terms of $\Omega$ as
\begin{equation}
  \begin{split}
 F^{(0)} &\simeq \frac{TV}{4\pi^2} \int_0^\infty dp_\perp p_\perp \int_{-\infty}^\infty dp^z \\
 &\quad\times 
 \left( 1 + \frac{\Omega^2 p_\perp^2 R^2}{8} \frac{d^2}{d\varepsilon^2} \right) \ln(1 - e^{-\beta \varepsilon})\,.
  \end{split}
\end{equation}
The first term in the parentheses simply corresponds to the negative pressure, $-p$, multiplied by $V$.  For the second term, we take the derivative with respect to $\varepsilon$ once, and then perform the integration by parts with respect to $p_\perp$, using $d/d\varepsilon=(\varepsilon/p_\perp)d/dp_\perp$.  After all, we find that the integrand takes the same structure as the second term in Eq.~\eqref{eq:expanded_F0}, which is proportional to the enthalpy density.  Thus, we finally reach
\begin{equation}
 F^{(0)} = -V p - V \frac{\Omega^2 R^2}{4} h + O(\Omega^4)\,.
\end{equation}
From this, the moment of inertia in Eq.~\eqref{eq:IR} can be readily reproduced.

We now compare the results obtained in this paper with those reported in Ref.~\cite{Siri:2024scq}.
Let us quote the free energy for the noninteracting gas from Eq.~(3.15) of Ref.~\cite{Siri:2024scq}, taking $x = m \beta \to 0$ and keeping only terms up to $\Omega^2$, as
\begin{equation}
 \frac{F^{(0)}_{\text{Ref.~\cite{Siri:2024scq}}}}{g V} = -\frac{\pi^2 T^4}{90} - \frac{T^2 \Omega^2}{12} \sum_{\ell = -\infty}^\infty \ell^2\,,
\end{equation}
where we also divided by a degeneracy factor of $g = 2$ taking into account the degrees of freedom of the charged scalar field considered in Ref.~\cite{Siri:2024scq}. It is clear that the summation over $\ell$ requires a regularization that may depend on $p_\perp$ in principle.  The authors of Ref.~\cite{Siri:2024scq} assumed that the $\ell=\pm 1$ contributions are dominant in the summation, setting
$\sum_{\ell = -\infty}^\infty \ell^2 \to 2$, as evident from Eq.~(3.17) therein. This leads to the erroneous result for the moment of inertia of the non-interacting theory [see Eq.~(4.14)]:
\begin{equation}
 \frac{I^{(0)}_{{\rm Ref}.~\cite{Siri:2024scq}}}{g V} = \frac{T^2}{3},
\end{equation}
which is, strikingly, independent of the system's radial extent, in contradiction even with the classical result in Eq.~\eqref{eq:rkt:I}. We are lead to conclude that the main error in the calculation presented in Ref.~\cite{Siri:2024scq} is the incorrect treatment of the combination of the summation over $\ell$ and the spatial integration of the Bessel functions.

\section{Conclusion}\label{sec:conc}

In this paper, we computed the moment of inertia for a weakly interacting, neutral scalar field in the weak-rotation limit. We establish the result in Eq.~\eqref{eq:claim}, namely that the moment of inertia density $\mathcal{I}$ is equal to the product between the enthalpy density $h$ and the squared distance to the rotation axis, $r_\perp^2$. 

We first established our claim via a kinetic theory analysis of the free Bose gas under rotation. We then provided a heuristic derivation starting from the free energy of the rotating scalar field, expressed using cylindrical modes. We showed that the improper use of the integration over the entire space, which eliminates the coordinate dependence appearing through the Bessel functions, followed by the infinite summation over the angular quantum number $\ell$, as performed in Ref.~\cite{Siri:2024scq}, leads to an inconsistent result. 

We finalized our analysis by first expressing the moment of inertia in the static limit in terms of two- and four-point functions, which we obtained using standard thermal field theory techniques, developed for static systems. Performing the calculations in perturbation theory up to $O(\lambda^{3/2})$ with respect to the self-interaction coupling strength $\lambda$, we confirmed our claim at arbitrary field vacuum mass $m$.

Our result shows that self-interactions do indeed reduce the moment of inertia, as a consequence of the increase of the thermal mass and a subsequent decrease of the enthalpy density. However, since $h > 0$, self-interactions cannot account for the emergence of a negative moment of inertia. As further avenues of research, we mention including higher-order corrections in perturbation theory, or the use of the functional renormalization group (FRG) technique to access the strong coupling limit, as discussed in Ref.~\cite{Chen:2023cjt}; 
computing the moment of inertia in the fast rotation regime, using cylindrical modes within a bounded system, as described in Ref.~\cite{Kuboniwa:2025vpg}; or extending the analysis to other quantum fields, such as the Dirac, electromagnetic and gluon fields, or the linear sigma model with quarks.


\begin{acknowledgments}
AG and VEA gratefully acknowledge support via EU’s NextGenerationEU instrument, through the National Recovery and Resilience Plan of Romania - Pillar III-C9-I8, managed by the Ministry of Research, Innovation and Digitization, within the project entitled ``Facets of Rotating Quark-Gluon Plasma'' (FORQ), contract no.~760079/23.05.2023 code CF 103/15.11.2022.
KF is supported in part by Japan Society for the Promotion of Science (JSPS) KAKENHI Grant Nos.\ 22H05118, 25K24464, 26K00698.
KF gratefully acknowledges the Visiting@WUT programme, contract no.\ UVT/076901/24.10.2025.
\end{acknowledgments}

\appendix

\section{Useful formulas}\label{app:formulas}

\subsection{Matsubara frequency summation}\label{app:formulas:sumn}

The one-loop calculation frequently meets the following summations with respect to the Matsubara frequency, $\omega_n = 2\pi n T$, i.e., 
\begin{align}
  T \sum_n \frac{1}{\omega_n^2 + \omega^2} &= \frac{1}{\omega} \left[ \nb(\omega) + \frac{1}{2} \right] \,,  \label{eq:sumn_D}\\
  T \sum_n \frac{\omega_n^2}{\omega_n^2 + \omega^2} &= -\omega \left[\nb(\omega) + \frac{1}{2}\right] + C\,,  \label{eq:sumn_omegan}\\
  T \sum_n \frac{1}{(\omega_n^2 + \omega^2)^2} &= \frac{1}{2\omega^3} \left[ \nb(\omega) + \frac{1}{2} \right] - \frac{1}{2\omega^2} \frac{d\nb}{d\omega}
 \label{eq:sumn_dD}
\end{align}
with $\nb(\omega) = (e^{\beta\omega} - 1)^{-1}$ the Bose-Einstein distribution function.
In the above expression, $C$ is an ultraviolet divergent constant; $C=T\sum_n$, which is independent of $\omega$, that is, $\propto \delta(\tau)$ in Euclidean time.  Thus, $C$ originates from the contact singularity, and we adopt a regularization to set $C=0$.
These expressions can be derived starting from the mixed representation of the Matsubara propagator \cite{Bellac:2011kqa},
\begin{equation}
 T \sum_n \frac{e^{-i \omega_n \tau}}{\omega_n^2 + \omega^2} = \frac{1}{2\omega} \left[(1 + \nb(\omega)) e^{-\omega \tau} + \nb(\omega) e^{\omega \tau}\right].
\end{equation}
The zeroth and second Maclaurin coefficients with respect to $\tau$ correspond to Eqs.~\eqref{eq:sumn_D} and \eqref{eq:sumn_omegan}, respectively.
Equation~\eqref{eq:sumn_dD} can be derived from Eq.~\eqref{eq:sumn_D} by applying the operator $-\frac{1}{2\omega} \frac{d}{d\omega}$ on both sides of the equal sign.

\subsection{High-temperature expansion}\label{app:formulas:highT}

The thermal expectation values considered in the main text may be expressed with respect to the following functions:
\begin{equation}
 h_n(y) = \frac{1}{\Gamma(n)} \int \frac{dx\, x^{n-1}}{\sqrt{x^2 + y^2}} \frac{1}{e^{\sqrt{x^2 + y^2}} - 1},
 \label{eq:hn_def}
\end{equation}
whose properties are thoroughly addressed in Ref.~\cite{Kapusta:2007xjq}. Knowing the small-$y$ expansion of $h_1(y)$, 
\begin{equation}
 h_1(y) = \frac{\pi}{2y} + \frac{1}{2} \ln \frac{y}{4\pi} + \frac{\gamma}{2} - \zeta(3) \left(\frac{y}{4\pi}\right)^2 + O(y^4),
\end{equation}
it is possible to obtain the small-$y$ expansion for the functions $h_{2n+1}(y)$ using the recurrence relation
\begin{equation}
 \frac{dh_{n+1}}{dy} =-\frac{y}{n} h_{n-1}(y).
\end{equation}
Knowing the integration constants $h_3(0)= \pi^2 / 12$ and $h_5(0) = \pi^4 / 360$, we obtain
\begin{subequations}\label{eq:h35}
\begin{align}
 h_3(y) &= \frac{\pi^2}{12} - \frac{\pi y}{4} + \frac{y^2}{8} \left(\frac{1}{2} - \gamma + \ln \frac{4\pi}{y}\right) + O(y^4), \label{eq:h3}\\
 h_5(y) &= \frac{\pi^4}{360} - \frac{\pi^2 y^2}{96} + \frac{\pi y^3}{48} + O(y^4). 
 \label{eq:h5}
\end{align}
\end{subequations}

\section{Self-energy under rotation} \label{App:Pi_Omega}

The purpose of this appendix is to validate the claim in Eq.~\eqref{eq:selfenergy_rot} regarding the dependence on the rotation angular velocity of the self-energy. To derive this expression, we solve the gap equation~\eqref{eq:Pi} self-consistently, using the cylindrical mode summation in Eq.~\eqref{eq:dP_cylindrical}:
\begin{equation}
 \Pi(\Omega) = \frac{\lambda}{2} \int_\ell \frac{d^3p}{(2\pi)^3 \varepsilon_\Pi} \nb(\tilde{\varepsilon}_\Pi)
 \label{eq:Pi_Omega}
\end{equation}
with $\nb(\omega) = 1/(e^{\beta \omega} - 1)$, $\tilde{\varepsilon}_\Pi = \varepsilon_\Pi - \ell \Omega$ and $\varepsilon_\Pi = \sqrt{\varepsilon^2 + \Pi(\Omega)}$.
We take the Maclaurin expansion of $\Pi$ with respect to the angular velocity to second order, i.e.,
\begin{equation}
 \Pi(\Omega) \simeq \Pi_0 + \Omega \Pi_1 + \frac{1}{2} \Omega^2 \Pi_2 \,,
 \label{eq:app_Pi_Omega_exp}
\end{equation}
where $\Pi_0$ satisfies 
\begin{equation}
 \Pi_0 = \frac{\lambda}{2} \int dP_0\, n_0\,, \qquad 
 dP_0 = \frac{d^3p}{(2\pi)^3 \varepsilon_0}\,,
 \label{eq:app_Pi0}
\end{equation}
where we use the zeroth-order notations as $n_0 = \nb(\varepsilon_0)$ and $\varepsilon_0 =\sqrt{\varepsilon^2 + \Pi_0}$, corresponding to the nonrotating case.
In this Appendix, the subscript of $\Pi_0, \Pi_1, \Pi_2, \dots$ refers to the expansion order with respect to $\Omega$.  The discussions in what follows below does not require any expansion with respect to $\lambda$, but we always assume that $\Pi$ is momentum independent.

We first show $\Pi_1=0$. Considering the expansion of the integrand to first order with respect to $\Omega$, we have
\begin{align}
 \frac{1}{\varepsilon_\Pi} &\simeq \frac{1}{\varepsilon_0} - \frac{\Omega \Pi_1}{2\varepsilon_0^3}\,, &
 \nb(\tilde{\varepsilon}_\Pi) &\simeq n_0 + \Omega \left(\frac{\Pi_1}{2\varepsilon_0} - \ell\right) \frac{dn_0}{d\varepsilon_0}\,.
\end{align}
Plugging this into Eq.~\eqref{eq:Pi_Omega} gives the following relation for $\Pi_1$:
\begin{multline}
 \Pi_1 \left[1 + \frac{\lambda}{4} \int \frac{dP_0}{\varepsilon_0} \left(\frac{1}{\varepsilon_0} - \frac{d}{d\varepsilon_0}\right) n_0 \right] \\
 = -\frac{\lambda}{2} \int dP_0 \frac{dn_0}{d\varepsilon_0} \sum_{\ell = -\infty}^\infty \ell [J_\ell(p_\perp r_\perp)]^2 \,,
\end{multline}
where we replaced $dp_\perp p_\perp dp^z = 4\pi^2 \varepsilon_0 dP_0$. Since the sum over $\ell$ vanishes by virtue of Eq.~\eqref{eq:sum_ell}, we conclude that $\Pi_1 = 0$.

We now expand the integrand up to $O(\Omega^2)$ using $\Pi_1=0$ via
\begin{align}
 \frac{1}{\varepsilon_\Pi} &\simeq \frac{1}{\varepsilon_0} - \frac{\Omega^2 \Pi_2}{4\varepsilon^3_0}\,, \nonumber\\
 \nb(\tilde{\varepsilon}_\Pi) &\simeq n_0 - \ell \Omega \frac{dn_0}{d\varepsilon_0} + \frac{\Omega^2}{2} \left(\ell^2 \frac{d^2n_0}{d\varepsilon_0^2} + \frac{\Pi_2}{2\varepsilon_0} \frac{dn_0}{d\varepsilon_0}\right).
 \label{eq:nrotexp}
\end{align}
Using Eqs.~\eqref{eq:sum_ell} to perform the summation over $\ell$, we arrive at the following equation for $\Pi_2$:
\begin{multline}
 \Pi_2 \left[1 + \frac{\lambda}{4} \int dP_0 \left( \frac{1}{\varepsilon_0^2} - \frac{1}{\varepsilon_0} \frac{d}{d\varepsilon_0}\right) n_0\right] \\
 = \frac{\lambda r_\perp^2}{6} \int dP_0\, \bp^2 \frac{d^2n_0}{d\varepsilon_0^2} \,.
\end{multline}
Employing integration by parts in both $dP_0$ integrals and taking the massless limit $m = 0$ gives
\begin{equation}
 \Pi_2 \left(1 + \frac{\lambda}{4} \int \frac{dP_0}{\bp^2} n_0 \right)
 = 2 r_\perp^2  \left(\Pi_0 + \frac{\lambda \Pi_0}{4} \int \frac{dP_0}{\bp^2} n_0\right),
\end{equation}
where we used the expression of $\Pi_0$ in Eq.~\eqref{eq:app_Pi0} on the right-hand side. 
Since we can drop the same nonzero factor from both sides,
we conclude $\Pi_2\rvert_{m = 0} = 2r_\perp^2 \Pi_0$. Substituting this into Eq.~\eqref{eq:app_Pi_Omega_exp}, we arrive at
\begin{equation}
 \Pi(\Omega)\rvert_{m = 0} \simeq \Pi_0(1 + r_\perp^2 \Omega^2)\,,
\end{equation}
thereby we establish Eq.~\eqref{eq:selfenergy_rot}. 

\section{Ring diagrams and the gap equation} \label{app:Pi}

In this Appendix, we briefly present the consistency between the self-consistent mass gap equation shown in Eq.~\eqref{eq:Pi} and the conventional description of the ring diagram resummation.

At first order in perturbation theory, the self-energy is given by the Hartree-type diagram shown in Eq.~\eqref{eq:Pi_tadpole},
with the dressed propagator replaced by the free one:
\begin{align}
  \Pi^{(1)} &= - \begin{tikzpicture}[baseline={([yshift=-.5ex]z)}]
    \begin{feynman}[small]
      \vertex [] (x) {};
      \vertex [right=of x, dot] (z) {};
      \vertex [right=of z] (y) {};
      \diagram* {
        (x) -- [plain] (z) -- [plain] (y),
      };
      \draw (z) arc [start angle=-90, end angle=270, radius=0.3cm];
    \end{feynman}
  \end{tikzpicture} - \begin{tikzpicture}[baseline={([yshift=-.5ex]z)}]
    \begin{feynman}[small]
      \vertex [] (x) {};
      \vertex [right=of x, crossed dot] (z) {};
      \vertex [right=of z] (y) {};
      \diagram* {
        (x) -- [plain] (z) -- [plain] (y),
      };      
    \end{feynman}
  \end{tikzpicture} \nonumber \\
  &= \frac{\lambda}{2} \propx_0(0) + \delta_m \,.
  \label{eq:Pi_tadpole_0}
\end{align}
Identifying $\delta_m = -\frac{\lambda}{2} G_{\rm vac}(0) = -\frac{\lambda}{4} \int dP$, the first-order approximation to the self-energy becomes $\Pi^{(1)} = \frac{\lambda}{2} \int dP\, \nb(\varepsilon)$.

The nonanalytic $\lambda^{3/2}$ contribution to the self-energy can be obtained by dressing the propagator in Eq.~\eqref{eq:Pi_tadpole_0} with extra self-energy bubbles, including the mass counterterm:
\begin{multline}
 \Pi^{(3/2)} = \Pi^{(1)} \\ +
 \begin{tikzpicture}[baseline={([yshift=-.5ex]z)}]
     \begin{feynman}[small]
       \vertex [] (x) {};
       \vertex [right=1cm of x, dot] (z) {};
       \vertex [right=1cm of z] (y) {};
       \vertex [above=0.8cm of z, empty dot] (z1) 
       {\,\tiny$\Pi^{\scalebox{0.6}{(1)} }$};
       \diagram* {
         (x) -- [plain] (z) -- [plain] (y),
       };
       \draw [plain] (z) arc [start angle=-90, end angle=58, radius=0.4cm];
       \draw [plain] (z) arc [start angle=270, end angle=122, radius=0.4cm];
     \end{feynman}
   \end{tikzpicture}  - \begin{tikzpicture}[baseline={([yshift=-.5ex]z)}]
    \begin{feynman}[small]
       \vertex [] (x) {};
       \vertex [right=1cm of x, dot] (z) {};
       \vertex [right=1cm of z] (y) {};
       \vertex [left=0.4cm of z] (a1) {};
       \vertex [right=0.4cm of z] (a2) {};
       \vertex [above=0.8cm of z] (zp) {};
       \vertex [above=0.4cm of a1, empty dot] (z1) 
       {\,\tiny$\Pi^{\scalebox{0.6}{(1)} }$};
      \vertex [above=0.4cm of a2, empty dot] (z1) 
      {\,\tiny$\Pi^{\scalebox{0.6}{(1)} }$};
          \diagram* {
            (x) -- [plain] (z) -- [plain] (y),
          };
          \draw [plain] (z) arc [start angle=-90, end angle=-33, radius=0.4cm];
          \draw [plain] (z) arc [start angle=270, end angle=213, radius=0.4cm];
          \draw [plain] (zp) arc [start angle=90, end angle=33, radius=0.4cm];
          \draw [plain] (zp) arc [start angle=90, end angle=147, radius=0.4cm];
        \end{feynman}
      \end{tikzpicture}
+ \begin{tikzpicture}[baseline={([yshift=-.5ex]z)}]
    \begin{feynman}[small]
       \vertex [] (x) {};
       \vertex [right=1cm of x, dot] (z) {};
       \vertex [right=1cm of z] (y) {};
       \vertex [left=0.5cm of z] (a1) {};
       \vertex [right=0.5cm of z] (a2) {};
       \vertex [above=0.5cm of z] (zp) {};
       \vertex [above=0.5cm of a1, empty dot] (z1) 
       {\,\tiny$\Pi^{\scalebox{0.6}{(1)} }$};
      \vertex [above=0.5cm of a2, empty dot] (z2) 
      {\,\tiny$\Pi^{\scalebox{0.6}{(1)} }$};
      \vertex [above=0.5cm of zp, empty dot] (z3) 
      {\,\tiny$\Pi^{\scalebox{0.6}{(1)} }$};
          \diagram* {
            (x) -- [plain] (z) -- [plain] (y),
          };
  \draw[plain] ([shift={(28:0.5cm)}]zp) arc [start angle=28, end angle=62, radius=0.5cm];
  \draw[plain] ([shift={(118:0.5cm)}]zp) arc [start angle=118, end angle=152, radius=0.5cm];
  \draw[plain] ([shift={(208:0.5cm)}]zp) arc [start angle=208, end angle=332, radius=0.5cm];
        \end{feynman}
      \end{tikzpicture} \\
      + \dots,
      \label{eq:Pi32_from_Pi1}
\end{multline}
which leads to 
\begin{equation}
 \Pi^{(3/2)} = \Pi^{(1)} - \frac{\lambda \Pi^{(1)}}{2} \sumint_{\:P} \widetilde{G}_0^2(P) \sum_{k = 0}^\infty [- \widetilde{G}_0(P) \Pi^{(1)}]^k\,.
\end{equation}
Performing the summation over $k$ leads to 
\begin{equation}
 \Pi^{(3/2)} = \frac{\lambda}{2} \sumint_{\:P} \widetilde{G}_1(P) + \delta_m\,, \quad 
 \widetilde{G}_1(P) = \frac{1}{\omega_n^2 + \varepsilon_1^2}
\end{equation}
with $\varepsilon_1^2 = \bm{p}^2 + m_1^2$ corresponding to the thermal mass $m_1^2 = m^2 + \Pi^{(1)}$. We define the regularized self-energy 
\begin{equation}
 \Pi_{\rm reg}^{(3/2)} = \frac{\lambda}{2} \sumint_{\:P} \widetilde{G}_1(P) + \delta_{m_1}
 = \frac{\lambda}{2} \int dP_1 \nb(\varepsilon_1),
\end{equation}
with $\delta_{m_1} = -\frac{\lambda}{4} \int dP_1$. Noting that $\Pi^{(3/2)}_{\rm reg} = \frac{\lambda}{2\pi^2} h_3(y_1)$ with $y_1 = \sqrt{m^2 + \Pi^{(1)}} / T$, the massless limit of $\Pi^{(3/2)}$ can be expressed using Eq.~\eqref{eq:h3} as
\begin{equation}
 \Pi_{\rm reg}^{(3/2)} \simeq \frac{\lambda T^2}{4!} - \frac{3T^2}{\pi} \left(\frac{\lambda}{4!}\right)^{3/2} + O(\lambda^2)\,.
 \label{eq:Pi32_small}
\end{equation}

The remainder $\delta_m - \delta_{m_1}$ due to the mismatch between the vacuum mass $m$ and the thermal mass $m_1$ leads to a logarithmic divergence, which appears though only at quadratic order in the coupling constant $\lambda$. Using the momentum cutoff regularization, we find
\begin{align}
 \delta_m - \delta_{m_1} &= \frac{\lambda}{8\pi^2} \int_0^\Lambda dp p^2 \left(\frac{1}{\varepsilon_1} - \frac{1}{\varepsilon_0}\right) \nonumber\\
 &\to \frac{\lambda \Pi^{(1)}}{16\pi^2} \ln\frac{e m}{2\Lambda}+ O(\lambda^3).
 \label{eq:Pi32_deltam}
\end{align}
This resummation process can be repeated by replacing $\Pi^{(1)}$ with the self-energy including $\Pi^{(3/2)}$ on the right-hand side of Eq.~\eqref{eq:Pi32_from_Pi1}, updating the self-energy. Continuing the process infinitely leads to the resummed solution, $\Pi_\mathrm{resum}$, of the gap equation in Eq.~\eqref{eq:Pi}.

\section{Large-coupling expansion}\label{app:largelambda}

Thermal expectation values in the large-coupling limit can be derived by taking into account that self-energy grows with the coupling $\lambda$. Thus, the large-coupling limit is equivalent to a high-mass expansion, or equivalently large $y$, with $y = \sqrt{m^2 + \Pi} / T$. Using the expansion of the distribution appearing in Eq.~\eqref{eq:hn_def} as
\begin{align}
\frac{1}{e^{\sqrt{x^2+y^2}}-1} = \sum_{\ell=1}^{\infty}e^{-\ell \sqrt{x^2+y^2}}
\end{align}
and subsequently substituting $z = \sqrt{x^2+y^2}/y$, one can show that Eq.~\eqref{eq:hn_def} can be expressed in terms of the modified Bessel functions $K_\nu(z)$ as
\begin{align}
h_{2 \nu+1}(y) = \frac{2^{\nu}\Gamma(\nu+\frac{1}{2})}{\sqrt{\pi}\Gamma(2 \nu+1)}y^{\nu} \sum_{\ell=1}^{\infty} \frac{K_{\nu}(\ell y)}{\ell^{\nu}}
\end{align}
Using the asymptotic expansion of the modified Bessel function of second kind for large arguments,
\begin{align}
K_{\nu} (z) = \sqrt{\frac{\pi}{2z}}e^{-z} [1-\frac{4 \nu^2-1}{8z}+O(z^{-2}) ],
\end{align}
and taking into account only $\ell=1$, we find for $\nu=1,2$:
\begin{align}
h_3(y) &= \frac{ \pi}{ 2}\left(\frac{y}{2\pi}\right)^{\frac{1}{2}}\left[1+\frac{3}{8y}+O(y^{-2})\right] e^{-y}, \\ 
h_5(y) &= \frac{ \pi^2}{ 4}(\frac{y}{2\pi})^{\frac{3}{2}}\left[1+\frac{15}{8y}+O(y^{-2})\right] e^{-y}.
\label{eq:largey}
\end{align}

Using the expression \eqref{eq:Pi} for the self energy $\Pi$ and the above high-$y$ expansion, we obtain 
\begin{align}
\Pi = \frac{\lambda T^2}{2} \frac{y^{1/2}}{(2 \pi)^{3/2}}e^{-y}. 
\label{eq:Pihighy}
\end{align}
This equation can be solved in the massless limit, $m = 0$, when $y = \sqrt{\Pi} / T$ and Eq.~\eqref{eq:Pihighy} becomes 
\begin{equation}
 e^y y^{3/2} = \frac{\lambda}{2(2\pi)^{3/2}}.
 \label{eq:Lambert_aux}
\end{equation}
The above equation can be solved in terms of the Lambert function, satisfying $e^{W(z)} W(z) = z$ \cite{DLMF,Corless:1996zz}. Raising this relation to the power $3/2$ gives
\begin{equation}
 e^{\frac{3}{2} W(z)} \left[\frac{3}{2} W(z)\right]^{3/2} = \left(\frac{3z}{2}\right)^{3/2},
\end{equation}
which indicates that Eq.~\eqref{eq:Lambert_aux} is solved by
\begin{equation}
 y = \frac{3}{2} W\left(\frac{\lambda^{2/3}}{3\pi \times 2^{2/3}}\right).
\end{equation}
This gives
\begin{align}
\Pi = \frac{9 T^2}{4} \left[W\left(\frac{\lambda ^{2/3} }{3\pi \times 2^{2/3}}\right)\right]^2.
\end{align}
Knowing the large-argument expansion of the Lambert function, $W(z) = \ln(z) - \ln[\ln(z)] + O(z^0)$ \cite{DLMF}, we obtain 
\begin{equation}
 \Pi \simeq \frac{9 T^2}{4} \left[\ln \left(\frac{\lambda^{3/2}}{3\pi \times 2^{2/3}}\right)\right]^2,
 \label{eq:largey_Pi}
\end{equation}
which shows that $\Pi$ grows with the square of the logarithm of $\lambda$.

Using the large-$y$ expansion \eqref{eq:largey} for the enthalpy in Eq.~\eqref{eq:rkt:h}, we find that
\begin{align}
h(y) = T^4 \left(\frac{y}{2 \pi}\right)^{3/2} y\, e^{-y} \left[1+\frac{35}{8y}+O(y^{-2})\right].
\end{align}
Additionally, taking into account Eq.~\eqref{eq:Pihighy}, previous relation can also be expressed as
\begin{align}
h(y) = \frac{2 \Pi^2}{\lambda} \left(1 + \frac{m^2}{\Pi^2}\right) \left[1+\frac{35}{8y} + O(y^{-2})\right].
\end{align}
At large coupling, where the mass $m$ plays a subleading role, Eq.~\eqref{eq:largey_Pi} can be used to show that 
\begin{equation}
 h \simeq \frac{81 T^4}{8\lambda} \left[\ln \left(\frac{\lambda^{3/2}}{3\pi \times 2^{3/2}}\right)\right]^4.
 \label{eq:enthlarge}
\end{equation}
The above result indicates that the enthalpy density $h$ decreases in the strong-coupling limit.

\bibliography{bibl}




    \label{fig:placeholder}

\end{document}